\documentclass{article}

\usepackage[authoryear,longnamesfirst]{natbib}
\usepackage{setspace}
\usepackage{datetime}
\usepackage{fancyhdr}
\usepackage{tikz}
\usepackage{natbib}
\usepackage{float}
\usepackage{booktabs}
\usepackage{multirow}
\usepackage{subfig}
\usepackage{graphicx}
\usepackage{amssymb,amsmath,amsthm}
\def\tsc#1{\csdef{#1}{\textsc{\lowercase{#1}}\xspace}}
\tsc{WGM}
\tsc{QE}
\tsc{EP}
\tsc{PMS}
\tsc{BEC}
\tsc{DE}
\RequirePackage{colortbl} % To provide color for soul (for english editing), for adding cell color of table
\RequirePackage{soul} % To highlight text

\newcommand{\COM}[1]{}

\usepackage{hyperref}
\hypersetup{
colorlinks=true,
linkcolor=blue,
anchorcolor=blue,
citecolor=blue,
filecolor=blue,      
urlcolor=blue,}

\usepackage{PRIMEarxiv}

\usepackage[authoryear,longnamesfirst]{natbib}
\usepackage{setspace}
\usepackage{datetime}
\usepackage{fancyhdr}
\usepackage{tikz}
\usepackage{natbib}
\usepackage{float}
\usepackage{booktabs}
\usepackage{multirow}
\usepackage{subfig}
\usepackage{graphicx}
\usepackage{amssymb,amsmath,amsthm}

\title{Spatio-temporal Joint Modelling on Moderate-Extreme Air Pollution in Spain}

\author{
  Kai Wang \\
  Wisdom Lake Academy of Pharmacy \\
  Xi'an Jiaotong-Liverpool Univerisity \\
  Department of Mathematical Sciences\\
  University of Liverpool\\
  \texttt{Kai.Wang17@student.xjtlu.edu.cn} \\
   \And
  Chengxiu Ling \\
  Wisdom Lake Academy of Pharmacy \\
  Xi'an Jiaotong-Liverpool Univerisity \\
  \texttt{Chengxiu.Ling@xjtlu.edu.cn} \\
     \And
  Ying Chen \\
  Wisdom Lake Academy of Pharmacy \\
  Xi'an Jiaotong-Liverpool Univerisity \\
  \texttt{Ying.Chen01@xjtlu.edu.cn} \\
     \And
  Zhengjun Zhang \\
  Department of Statistics\\
   University of Wisconsin–Madson \\
   School of Economics and Management\\
   University of Chinese Academy of Sciences\\
  \texttt{zjz@stat.wisc.edu} \\
}

\begin{document}
\maketitle

\begin{abstract}
Very unhealthy air quality is consistently connected with numerous diseases. Appropriate extreme analysis and accurate predictions are in rising demand for exploring potential linked causes and for providing suggestions for the environmental agency in public policy strategy. This paper aims to model the spatial and temporal pattern of both moderate and extremely poor {PM}$_{10}$ concentrations (of daily mean) collected from 342 representative monitors distributed throughout mainland Spain from 2017 to 2021. 
We firstly propose and compare a series of Bayesian hierarchical generalized extreme models of annual maxima {PM}$_{10}$ concentrations, 
including both the fixed effect of altitude, temperature, precipitation, vapour pressure and population density, as well as the spatio-temporal random effect with the Stochastic Partial Differential Equation (SPDE) approach and a lag-one dynamic auto-regressive component (AR(1)).  Under WAIC, DIC and other criteria, the best model is selected with good predictive ability based on the first four-year data (2017--2020) for training and the last-year data (2021) for testing. 
We bring the structure of the best model to establish the joint Bayesian model of annual mean and annual maxima {PM}$_{10}$ concentrations and provide evidence that certain predictors (precipitation, vapour pressure and population density) influence comparably while the other predictors (altitude and temperature) impact reversely in the different scaled {PM}$_{10}$ concentrations. The findings are applied to identify the hot-spot regions with  poor air quality using excursion functions specified at the grid level. It suggests that the community of Madrid and some sites in northwestern and southern Spain are likely to be exposed to severe air pollution, simultaneously exceeding the warning risk threshold. 

\end{abstract}

% keywords can be removed

\footnote [1]{This project is supported by the Research Development Fund at XJTLU (RDF1912017), 
the Post-graduate Research Scholarship (PGRS2112022) and Jiangsu Qinglan Talent in 2022.}
\footnote [2]{Corresponding author: Chengxiu Ling, email: Chengxiu.Ling@xjtlu.edu.cn}

\section{Introduction}

Air pollution, composed of particulate matter ($\mathrm{PM}$) and gaseous pollutants, has a substantial negative impact on the environment, ecosystem and human health. Poor air quality is one of the five most significant health risks worldwide, alongside high blood pressure, smoking, diabetes and obesity \citep{COHEN20171907, SourceofPM}. It becomes one of the most considerable health concerns for the residents in areas of higher population density \citep{AP2}, centres with dense activities, and to particular user groups \citep{AP3, AP4}. Among all the pollutants, the particulate matter, with an aerodynamic diameter less than or equal to 10 and 2.5 microns respectively ($\mathrm{PM}_{10}$ and $\mathrm{PM}_{2.5}$) are most consistently connected with numerous adverse health outcomes including lung infections, cardiovascular diseases, and respiratory problems \citep{pm10, Martuzzi,  medparticle}, while appropriate regulation directly reduces the adverse health effects, increases general well-being, and improves public health \citep{STEINLE2015383}.

In Europe, although the European Environment Agency (EEA) maintains a rather dense particulate matter monitoring network to record the concentration levels across countries, huge regions of the European continent remain unmonitored. For proper assessment of population-wide exposure and appropriate formulation of pollution mitigation strategies, the responsible authorities need accurately estimate and predict the concentration levels at the unobserved locations \citep{CHU2015176}. 

The main challenge to forecasting particulate matter concentrations corresponds to the complexity of PM generation and spreading dynamics. On the one hand, the PM generation is dominated by two complicated sources, inorganic aerosol coming from the agriculture, long-range transport and energy sectors \citep{STEINLE2013184, SourceofPM}, as well as organic aerosol coming from biomass and fossil fuels burning emissions, vehicles emissions and cooking \citep{AP6, AP5}. On the other hand, the PM spread depends on both meteorological conditions and land use dispersion, leading the observed concentration levels to fluctuate geographically and temporally. 

In the statistical literature, the Bayesian spatio-temporal model that allows modelling a complex environmental phenomenon through a hierarchy of sub-models becomes one of the most promising methodologies in air quality scientific investigations \citep{Cameletti2013, Extremepm10, Taheri, Forlani, Guido, bgev}. In particular, this approach allows involving the explanatory variables to explain the large-scale variability, take residual dependency into account through a space-time process with a Gaussian {r}andom {f}ield (GRF), and produce high-resolution spatial forecasts to meet the rising demand for predictive concentration maps in epidemiological studies.

According to \cite{Porcu}, the main drawback of the Bayesian model with a GRF %Gaussian Random Field (GRF) 
refers to the computational difficulty to deal with enormous amounts of data, especially applying complex spatial dependence measures (i.e., the Mat\'{e}rn covariance function). Some strategies have been proposed to alleviate the computational burden of fitting complex spatial and temporal models. \cite{L1} proposed the {s}tochastic {p}artial {d}ifferential {e}quation (SPDE) approach, providing a method to represent a continuous Mat\'{e}rn field  through a discretely indexed Gaussian Markov random field (GMRF) associated with a sparse precision matrix, which enjoys good computational property. \cite{Rue1} also provided the Integrated Nested Laplace Approximation (INLA) algorithm that performs direct numerical calculations on the marginal posterior distributions, avoiding the time-consuming Markov chain Monte Carlo (MCMC) simulations. Additionally, GMRF with SPDE approach can be fitted in a Bayesian hierarchical framework through the INLA approach, with implementation in the R-INLA package  available at \url{https://www.r-inla.org/}, making this methodology fast and easily implemented.

Most previous spatial and temporal studies on air pollution only concentrated on moderate (i.e., daily, monthly and annual mean)     %moderate 
PM concentrations \citep{Cameletti2013, Beloconi2018BayesianGM, Guido, SAEZ2022105369}. However, extreme conditions are actually more concerned with environmental quality management due to their various hazardous impacts \citep{Extremepm10}. Numerous epidemiological studies pointed out that short-term exposures to severe particulate matter pollution can trigger serious acute {cardiovascular and respiratory mortality \citep{ORELLANO2020,LEI2019,ZHANG2019,shortterm,BJextreme}} and huge economic loss in the corresponding hospitalization {\citep{XIE2021, SHAH20131039}}. In the field of extreme case spatio-temporal analysis, {\cite{SHARMA2012170, RODRIGUEZ2016,Extremepm10, MARTINS201744} and \cite{bgev}} typically focused on a small spatial domain, making it difficult to consider the complicated orography with a variety of climatic conditions, as well as to provide general suggestions to national governments on the environmental policy formulation and health care allocation. More importantly, to our best knowledge, no studies consider the potential difference between moderate and extreme air pollution, in other words, model different scaled air pollution simultaneously to identify similarities and differences in the effects of influential factors.

In this paper, we focus on the spatial and temporal variation of both moderate and extreme air pollution  (i.e., annual mean and annual maxima of daily $\mathrm{PM}_{10}$ concentration levels) in Spain from 2017 to 2021, {after controlling for meteorological variables and socio-economic factors}. {The contribution is two-fold, the predictions of extreme pollution (excursion functions) and the investigation of similar/reverse effects of predictors in different scaled cases. Firstly, we establish several Bayesian hierarchical generalized extreme models on annual maxima and select the best model based on their predictive performance, detailed in Sections \ref{EVTmodel} and \ref{compare}. {Secondly, we utilize the joint Bayesian model with sharing effects in Sections \ref{joint} and \ref{Joint output} to model both annual mean and maxima concentrations simultaneously. We observe the comparable influence of precipitation, vapour pressure, and population density, as well as the possible opposite effects of altitude and temperature. We also generate excursion function maps \citep{excursion} based on the joint model to highlight the regional risk ranking that simultaneously exceeds the warning risk threshold in Section \ref{Prediction}.}
 }
These main findings, comprehensive knowledge on the $\mathrm{PM}_{10}$ generation and spread with high-resolution spatial forecasts are expected to promote awareness of the significance of extreme air pollution research, help in the investigation of the long-term effect in epidemiological studies, and underpins   air pollutants regulation and human health protection strategies for environmental agencies.

The rest of the paper is organized as follows. In Section \ref{Data}, we demonstrate the dataset for response and main explanatory variables. In Section \ref{method}, we formalize spatio-temporal Bayesian generalized extreme models on moderate-extreme air pollution in the framework of Bayesian spatial analysis and extreme value theory. We present the main results, applications and potential influence in Section \ref{RandD}, and conclude this paper with extensional discussions in Section \ref{conclusion}.

\section{Data}\label{Data}
\subsection{{PM}$_{10}$ concentrations and spatial domain}
\begin{figure}[ht]
    \centering
    \includegraphics[width=0.6\textwidth]{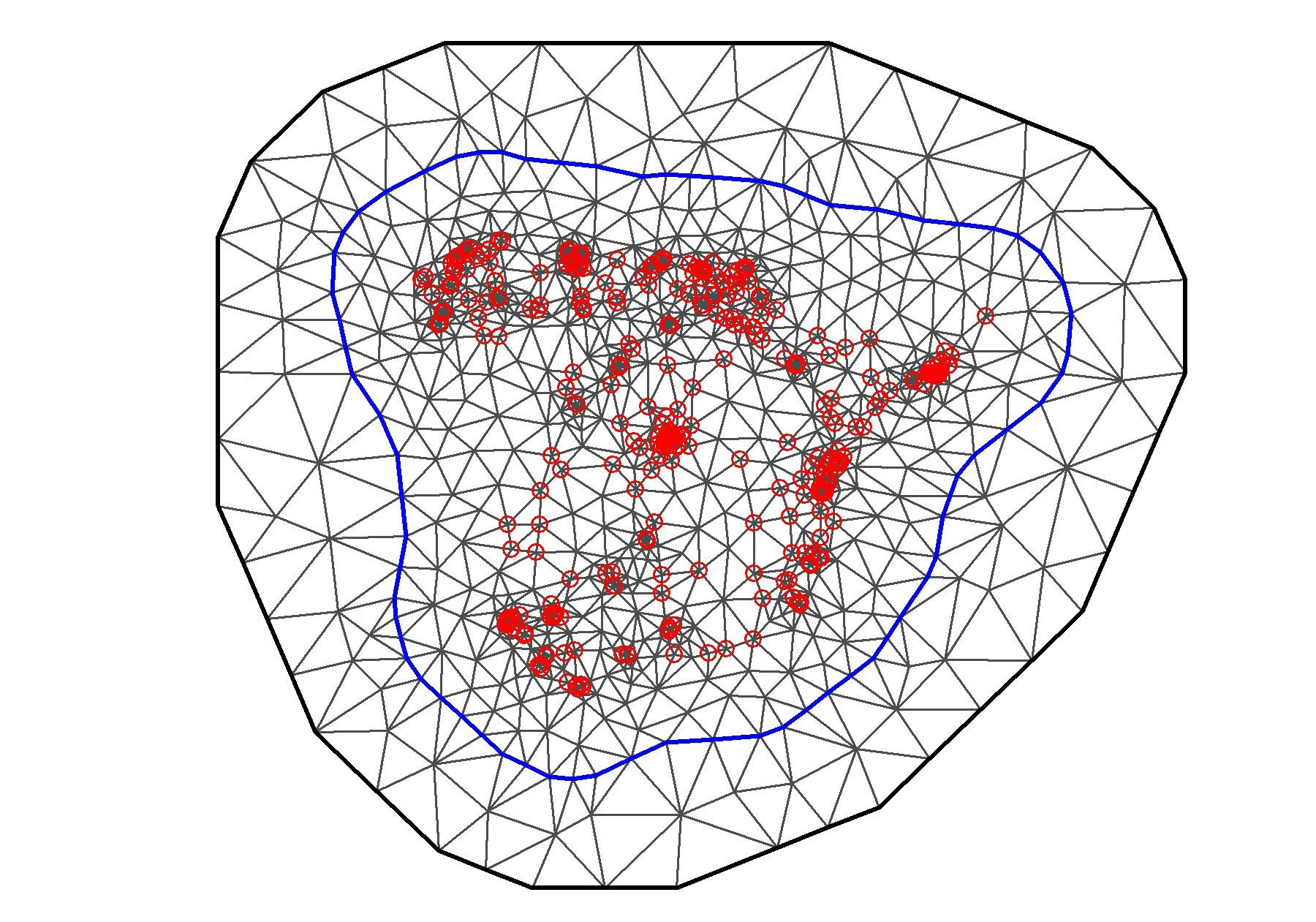}
    \caption{Study domain together with the spatial distribution of the 342 monitoring sites in \textcolor{red}{red} circles. The figure also illustrates the mesh used to build the SPDE approximation to the continuous Mat\'{e}rn field.}
    \label{Mesh}
\end{figure}

In mainland Spain, the $\mathrm{PM}_{10}$ concentration levels data is accessible by the Air Quality e-Reporting which is EEA’s air quality database consisting of a multi-annual time series data of air quality measurement 
and calculated statistics for a number of air pollutants. To work with a more robust dataset, we only retain 342 air pollution stations that have at least 60\% valid observations in a year, with the geographical distribution of the stations shown in red circles embedded in the mesh constructed for the SPDE approach (Figure \ref{Mesh}). This five-year dataset (2017--2021) with 1470 observations is divided into the training set (2017--2020; 1215 observations) and the validation set (2021; 255 observations), which are used to evaluate and compare the model fitness and predictive ability.

\begin{figure}[htbp!]
    \centering
    \includegraphics[width=0.8\textwidth]{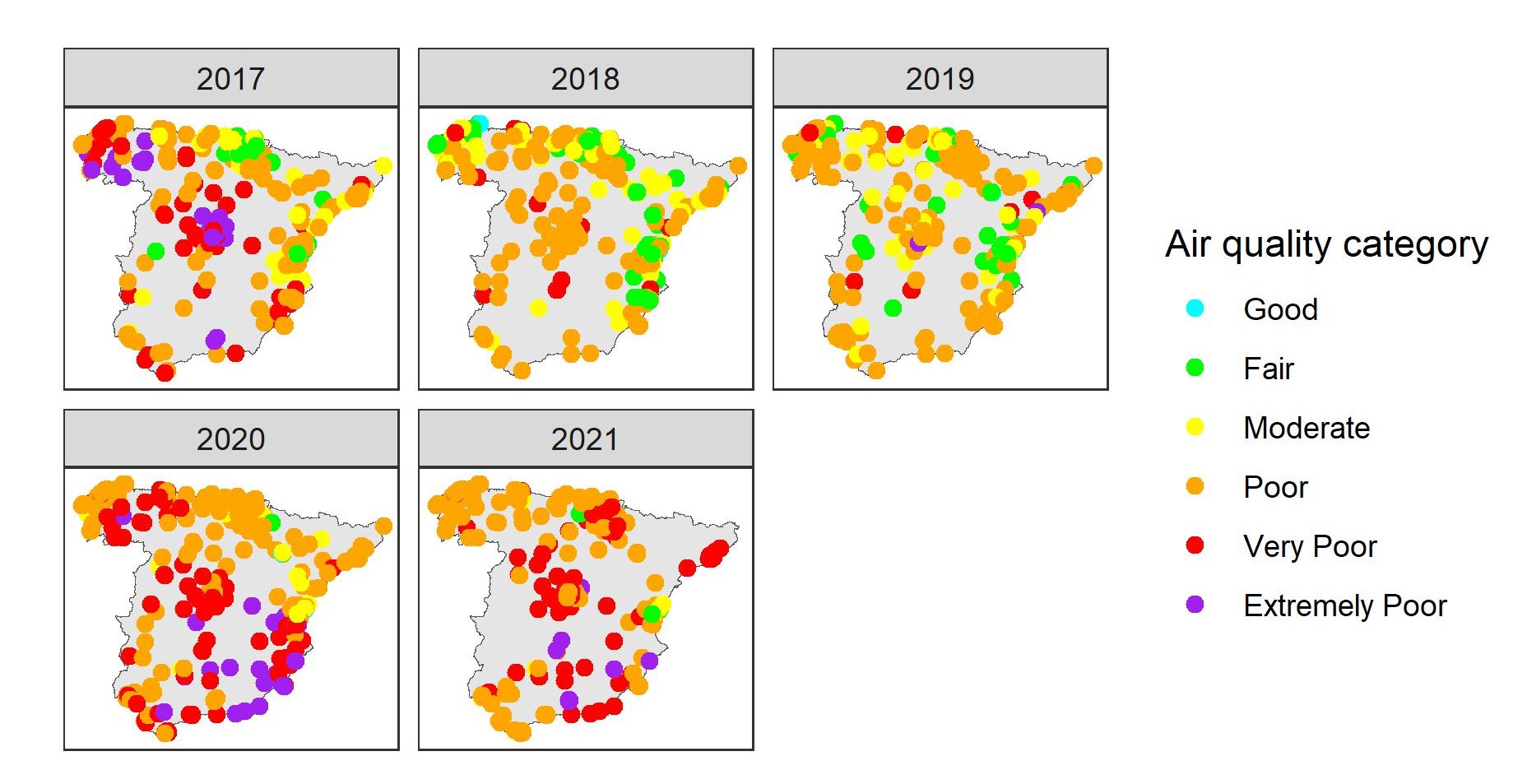}
    \caption{Spatio-temporal patterns of annual maxima $\mathrm{PM}_{10}$ concentration levels in {year} 2017--2021 throughout mainland Spain. The annual data is reported by European Environment Agency (EEA) and shown in the heat map with EEA's air quality category.}
        
    \label{pm10}
\end{figure}

%The temporal and spatial variations of extreme $\mathrm{PM}_{10}$ pollution are shown in detail in Figure \ref{pm10}. 
{Figure \ref{pm10} shows the temporal and spatial variations of extreme $\mathrm{PM}_{10}$ following the EEA's air quality categories with 0--20$\mu \mathrm{g} / \mathrm{m}^3$ (Good), 20--40$\mu \mathrm{g} / \mathrm{m}^3$ (Fair),  40--50$\mu \mathrm{g} / \mathrm{m}^3$ (Moderate), 50--100$\mu \mathrm{g} / \mathrm{m}^3$ (Poor), 100--150$\mu \mathrm{g} / \mathrm{m}^3$ (Very poor),  more than 150$\mu \mathrm{g} / \mathrm{m}^3$ (Extremely poor).}
Temporally, severe pollution seems to occur in 2017 and 2020, as indicated by the numerous monitors coloured in red (very poor) and purple (extremely poor). Spatially, compared with relatively low annual maxima recorded in the east (Valencian Community) and north (Basque Country), high $\mathrm{PM}_{10}$ concentrations are most prevalent in the centre, northwest and southeast, which correspond to the autonomous communities of Madrid, Galicia, Andalusia, and the Region of Murcia, respectively. This spatio-temporal pattern inspires our further investigation of spatio-temporal modelling taking appropriate topography covariates into account, which are stated in the following section.

\subsection{Explanatory variables}\label{explanatory}
A number of potential predictors are available based on prior findings in the air quality literature \citep{Cameletti2013, Guido, bgev}, and we choose to include a set of five main spatial and spatio-temporal varying predictors with the complete description list reported in Table \ref{Summarycovarites}.

\begin{table}
\caption{Description for explanatory variables.}\label{Summarycovarites}
\setlength{\tabcolsep}{5mm}
\begin{tabular}{*{4}{llll}}
\toprule
Predictors  & Description & Units & Spatial Resolution  \\
\midrule
Altitude  & Altitude of $\mathrm{PM}_{10}$ monitors  & m & Station Specific \\
Temperature  & Annual mean temperature  & °C  & 50km $\times$ 50km \\
Precipitation  & Annual mean precipitation & mm/month  & 50km $\times$ 50km \\
Vapour Pressure  &  Annual mean vapour pressure & hPa & 50km $\times$ 50km \\
Population Density  & Population per unit area & per $\mathrm{km^2}$ & Autonomous Communities \\
\bottomrule
\end{tabular}
\end{table}

In the following, we describe the selected predictors in detail.

\emph{Meteorological variables.} The meteorological variables (temperature, precipitation and vapour pressure) of the monthly mean are collected from the CRU TS 
\citep[Climatic Research Unit gridded Time Series;][]{Datasource} dataset and aggregated to be annual mean. Accordingly, CRU TS was first published in 2000, using ADW (angular-distance weighting) to interpolate anomalies of monthly observations onto a 0.5° grid over land surfaces (excluding Antarctica) for observed and derived variables (mean, minimum and maximum temperatures, precipitation, vapour pressure, wet days and cloud cover) with no missing values in the defined domain.

\emph{Elevation.} The altitude data, height over sea level, matched with locations of all air pollution monitors, is accessible in the annual aggregated air quality values dataset provided by EEA, available at \url{https://discomap.eea.europa.eu/App/AirQualityStatistics/index.html}.

\emph{Population density.} The population densities are calculated in each autonomous community. The original data is collected from the statistics report (available at \url{https://stats.oecd.org/}) of the Organisation for Economic Co-operation and Development (OECD). The OECD statistics contain data and metadata for economic and education indexes of OECD countries and some selected non-member economies.

We see in Table \ref{corr} that the correlations of potential predictors to  extreme and average PM$_{10}$ display a similar or different direction, which also vary  year by year. This will be further investigated by our spatio-temporal generalized extreme model and joint model with sharing effects, see details in both Sections \ref{EVTmodel} and \ref{joint}.
{Furthermore,  considering the correlation between location variables and meteorological variables (e.g., latitude and temperature), we adjust the location variables (longitude and latitude) as covariates to investigate the impact of meteorological factors.}

\begin{table}[H]
\centering
\caption{Correlation between explanatory variables and annual mean and annual maximum PM$_{10}$ on log scale.}
\resizebox{\linewidth}{!}{ 
\begin{tabular}{*{5}{lrrrr}}
\toprule
  \multirow{2}*{Explanatory variables} & \multicolumn{6}{c}{Correlation to $\ln{\text{mean}}$ ($\ln{\text{max}}$) in each year (sample size)}\\
  \cmidrule(rr){2-7}
  & 2017 (294) & 2018 (296) & 2019 (300) &  2020 (325) & 2021 (255) & Total (1470) \\
  \midrule
  Altitude & $-0.34\  $ $ \ (0.39) $ & $ -0.39\ \  $ $(0.39)$ & $-0.40$ $(0.13)$ & $-0.31$ $(-0.24)$ & $-0.24$ $(-0.01)$ &  $-0.34\ \ $ $(0.32)$\\
  Temperature & $0.28$ $ (-0.02) $ & $ 0.28\ \ $ $(0.01)$ & $0.34$ $(0.15)$ & $0.31\ \ $ $(0.19)$ & \ \ $0.33$ $(-0.13)$ &  $0.30\ \ $ $(0.08)$\\
  Precipitation & $\ -0.13$ $ (-0.19) $ & $ -0.08 $ $(-0.15)$ & $-0.14$ $(0.01)$ & $-0.21$ $(-0.27)$ & $-0.25$ $(-0.13)$ &  $-0.18$ $(-0.20)$\\
  Vapour Pressure & $0.28 $ $ (-0.15) $ & $ 0.33 $ $(-0.05)$ & $0.36$ $(0.17)$ & $0.32\ \ $ $(0.23)$ & \ \ $0.29$ $(-0.25)$ &  $0.31\ \ $ $(0.02)$\\
  Population Density & $0.06\ \ $ $ (0.28) $ & $ 0.04\ \  $ $(0.09)$ & $0.02$ $(0.02)$ & $0.01\ \ $ $(0.02)$ & $-0.01\ \ $ $(0.11)$ &  $-0.02\ \ $ $(0.10)$\\
  
  \bottomrule
\end{tabular}}
\label{corr}
\end{table}

\section{Model Formulation}\label{method} 
{
In Section \ref{EVTmodel}, we provide  first a brief introduction to spatio-temporal Gaussian field and extreme value models, followed by four candidate models for the annual maximum of daily PM$_{10}$ in mainland Spain. Subsequently, in Section \ref{joint}, we present the joint Bayesian Gumbel-Gaussian model for both extreme and moderate levels.}

\subsection{Spatio-temporal extreme value models with mixed effects}\label{EVTmodel}

{The generalized extreme value (GEV) distribution is widely employed for modelling extremes in environmental science \citep{Reiss2007}, such as temperature \citep{Cheng2014}, precipitation \citep{Panagoulia2014}, air pollution \citep{Zhang, MARTINS201744} and sea level \citep{Lobeto2018}. The GEV distribution has three parameters, location parameter ($\mu$, $-\infty<\mu<\infty$), scale parameter ($\sigma$, $\sigma>0$) and tail parameter ($\xi$, $-\infty<\xi<\infty$) with the cumulative distribution function 
$$
\mbox{GEV}(x; \mu,\sigma, \xi)=\exp \left\{-\left[1+\xi\left(\frac{x-\mu}{\sigma}\right)\right]_{+}^{-\frac{1}{\xi}}\right\}.
$$
%defined on the set $\{x: 1+\xi(x-\mu) / sigma>0\}$.
The case $\xi=0$ is interpreted as the limit case of $\xi \rightarrow 0$, leading to the Gumbel family with distribution function
$$
\mbox{Gumbel}(x; \mu, \sigma)=\exp \left\{-\exp \left[-\left(\frac{x-\mu}{\sigma}\right)\right]\right\}, \quad x \in \mathbb{R}.
$$}
{Considering the potential spatial and temporal dependence among the PM$_{10}$ concentrations, we use two typical approaches for measurement:  Mat\'{e}rn covariance function \citep{Mat, GG} with the Stochastic Partial Differential Equations approach \citep[SPDE; ][]{L1} approximation for spatial correlation and the auto-regressive dynamic model \citep[AR;][]{AR1model} for temporal dependence. We establish two groups of generalised extreme value models with similar fixed effects and varying random effects to model the annual maxima $\mathrm{PM}_{10}$ concentrations. }

\textbf{Model 1: Gumbel model with fixed effect and spatio-temporal random effect}. Let $y_{\text{max}}(\boldsymbol{s},t)$ denote the logarithm transform of annual maxima $\mathrm{PM}_{10}$ concentrations at location $\boldsymbol{s} \in \mathcal{S}$ and year $t \in \mathcal{T}$, where $\mathcal{S}$ is the study area and $\mathcal{T}$ is the time period in focus. Under the assumption of constant scale ($\sigma$) and tail ($\xi$) parameters, we use a linear combination of fixed effects with explanatory variables and spatio-temporal varying random effect to model the location parameter ($\mu(\boldsymbol{s},t)$) in Gumbel model  below.  Suppose that
\begin{equation}\label{Model 1}
\begin{split}
\left[y_{\text{max}}(\boldsymbol{s},t) \mid \mu(\boldsymbol{s},t), \sigma\right] & \sim \operatorname{Gumbel}\left(\mu(\boldsymbol{s},t), \sigma\right)\\
\mbox{with}  \quad
\mu(\boldsymbol{s},t)& = \mathbf{x}(\boldsymbol{s},t)^\top\boldsymbol{\beta}+u(\boldsymbol{s},t)
\end{split}
\end{equation}
and 
\begin{equation}\label{Spatio-temporal}
\begin{split}
u(\boldsymbol{s},t) & = a u(\boldsymbol{s},t-1 )
+w\left(\boldsymbol{s}, t\right), \\
w\left(\boldsymbol{s}, t\right) & \sim \mathcal{G} \mathcal{P}_{2 \mathrm{D}-\operatorname{SPDE}}\left(\rho_M, \sigma_M, \nu_M\right).
\end{split}
\end{equation}
Here, the fixed effects are associated with the vector $\mathbf{x}(s,t)$ including an intercept and the explanatory variables of location variables, meteorological variables and human-effect variables listed in Table \ref{Summarycovarites}, and the vector $\boldsymbol{\beta}$ corresponds to the regression coefficients associated with the fixed effects. {The term $u(\boldsymbol{s},t)$ represents a spatio-temporal varying random effect that incorporates spatio-temporal interaction \citep{Cameletti2013}. It temporally changes according to AR(1) dynamics with autocorrelation parameter $a$ and spatial correlated 
{and serially independent innovations } %effects 
$w\left(\boldsymbol{s}, t\right)$.} 

{Given two locations $s_i$ and $s_j$ separated by $h=d(s_i,s_j)$ (normally Euclidean) units, the Gaussian process $w(s,t)$ with mean 0 and Mat\'{e}rn covariance function is in the form of
\begin{eqnarray*}
\label{Matern}
\operatorname{Cov}(w(s_i,t),w(s_j,t'))= \begin{cases}0, & t \neq t' , \\ 
{\frac{\sigma^2}{2^{\nu-1} \Gamma(\nu)}\left(\sqrt{8 \nu} \frac{h}{\rho}\right)^{\nu} K_{\nu}\left(\sqrt{8 \nu} \frac{h}{\rho}\right) }, & t=t',\end{cases}
\end{eqnarray*}
where $\Gamma$ is the gamma function, $K_\nu$ is the modified Bessel function of the second kind,  $\rho>0$ is the range parameter, $\nu>0$ is the smoothness parameter, and $\sigma^2>0$ for the marginal variance.}

%The application of the Mat\'{e}rn covariance function in the latent Gaussian model rapidly increases after the proposal of the Stochastic Partial Differential Equations approach \citep[SPDE; ][]{L1}. Given a continuous Mat\'ern field defined on the domain in two-dimensional space (i.e., a second-order stationary and isotropic Gaussian Field $w(\boldsymbol s),\ \boldsymbol s\in\mathcal S\subset \mathbb{R}^2$ with a Mat\'{e}rn covariance function defined in Eq.\eqref{Matern}, denoted by $w(\boldsymbol{s})\sim \mathcal{GP}_{2\mathrm D-\operatorname{SPDE}}(\rho, \sigma, \nu)$, the SPDE approach basically uses a finite element representation to define the continuous Mat\'{e}rn field as a linear combination of basis functions defined on a triangulation of the domain. This representation makes it possible to use a Gaussian Markov random field (GMRF) as the best representative of the given Mat\'{e}rn field, while the GMRF with a local neighbourhood in a sparse precision matrix enjoys its good computational properties. This allows for avoiding the “big $n$ problem” that arises when working with the dense covariance matrix of the Gaussian Field \citep{Cameletti2013}. 

\textbf{Model 2: Gumbel model with fixed effect and separated spatial and temporal random effects.} Our second model is a modification of Model 1 specified in Eq.\eqref{Model 1} with {spatial and temporal random effects and {separable} interaction effects specified in Eq.\eqref{Spatio-temporal},} namely, we keep the Gumbel distribution assumption on $y_{\text{max}}(\boldsymbol s,t)$ as below. 
{
\begin{equation}
\label{Model 2}
  \begin{split}
  \left[y_{\text{max}}(\boldsymbol{s},t) \mid \mu(\boldsymbol{s},t), \sigma\right] & \sim \operatorname{Gumbel}\left(\mu(\boldsymbol{s},t), \sigma\right),\\
\mu(\boldsymbol{s},t)& = \mathbf{x}(\boldsymbol{s},t)^\top\boldsymbol{\beta}+f(t)+w(\boldsymbol{s}) + {u(\boldsymbol{s},t)},\\
\mbox{with}\ \      f(t) &\sim \mathcal{G} \mathcal{P}_{\mathrm{AR} (1)}\left(a, \tau_{AR}\right),\\
w\left(\boldsymbol{s}\right) & \sim \mathcal{G} \mathcal{P}_{2 \mathrm{D}-\operatorname{SPDE}}\left(\rho_M, \sigma_M, \nu_M\right).
    \end{split}
\end{equation}
}
{Note that $u(\boldsymbol{s},t)$ is defined the same as in Eq.\eqref{Spatio-temporal}, indicating the spatio-temporal interaction term with the Kronecker product, see e.g., \cite{Cameletti2011, Cameletti2013,Guido}.} The $f(t)$ denotes the non-linear random effect in the temporal structure of AR($1$), the $w(\boldsymbol{s}, t)$ is the spatially dependent only random effect with SPDE structure. Specifically, the implementation of the AR($1$) model in INLA generally assumes the Gaussian white noise with mean $0$ and precision $\tau_{AR}$.
For $f(t)$ defined over the naturally binned covariate (Year), let $\boldsymbol{t}=\left(t_1, \ldots, t_5\right)^\top$ denotes the time from the first year ($t_1$) to the last year ($t_5$), 
$$
\begin{aligned}
f(t_1) & \sim \mathcal{N}\left(0,\left(\tau_{AR}\left(1-a^2\right)\right)^{-1}\right), \\
f(t_i) &=a f(t_{i-1})+\epsilon_i , \quad \epsilon_i \sim \mathcal{N}\left(0, \tau_{AR}^{-1}\right), \quad i=2, \ldots, 5, 
\end{aligned}
$$
where $-1<a<1$ is a numeric constant, the so-called {autocorrelation}, by which we multiply the lagged variable $f(t_{i-1})$, and $\epsilon_i$ denotes the unpredictable error in the form of Gaussian white noise.

\textbf{Model 3: GEV model with fixed effect and spatio-temporal random effect.}
The generalised extreme value (GEV) models are basically following the same structure as Gumbel models except the generalized extreme value distribution for the response. To be specific, we suppose that 
\begin{equation}\label{Model 3}
\left[y_{\text{max}}(\boldsymbol{s},t) \mid \mu(\boldsymbol{s},t)\right] \sim \operatorname{GEV}\left(\mu(\boldsymbol{s},t), \sigma, \xi\right),
\end{equation}
where the location parameter $\mu(\boldsymbol{s},t)$ is of the same form of mixed effects as in Eq.\eqref{Model 1} and random effects in Eq. \eqref{Spatio-temporal}. 
%Note that this model generalizes the Gumbel model with spatio-temporal random effects, as the GEV model with $\xi=0$ corresponds to the Gumbel model.  

\textbf{Model 4: GEV model with fixed effect and separated spatial and temporal random effect}. Similar consideration of Model 2 modified from Model 1, we consider the following model in parallel with Model 3, i.e., we take  spatial and temporal random effects {with an  interaction effect} into the GEV model. Suppose that
\begin{equation}\label{Model 4}
    \begin{split}
     \left[y_{\text{max}}(\boldsymbol{s},t) \mid \mu(\boldsymbol{s},t), \sigma, \xi \right]  &\sim \operatorname{GEV}\left(\mu(\boldsymbol{s},t), \sigma, \xi \right)
    \end{split}
\end{equation}
with the location parameter $\mu(\boldsymbol{s},t)$ is of spatio-temporal structure of the form in Eq.\eqref{Model 2}. 

{It is worth pointing out that the well-known limit type theorem in \cite{Coles} motivates our GEV distribution assumption of  annual maxima of PM${_{10}}$, and its reduced case (i.e., Gumbel corresponds to GEV with $\xi=0$). The latter one becomes more parsimonious, and its exponential decay tail might be  appropriate to fit extreme air quality \citep{Zhang}. The Gumbel model is generally needed when the uncertainty of $\xi$ being non-zero arises in the GEV model with an estimate of $\xi$ close to zero.}

\subsection{Bayesian joint model with sharing effects}\label{joint}

In order to identify the potential varied effect levels of main explanatory variables, 
with inspiration from applications of the joint model with sharing effects on wildfire \citep{Koh2021SpatiotemporalWM}, we model both moderate and extreme PM$_{10}$ pollution simultaneously in two respective sub-models linked by the sharing effects and the sharing coefficients (scaling factors).

Let $y_{\text{mean}}(\boldsymbol{s},t)$ denote the logarithm transform of annual mean $\mathrm{PM}_{10}$ at location $\boldsymbol{s} \in \mathcal{S}$ and year $t \in \mathcal{T}$. We perform the Gaussian sub-model and Gumbel sub-model on annual mean and annual maxima {simultaneously, }
%, respectively, 
with the structure of the best extreme value model as Model 1 according to the model fitness and prediction analysed in Section \ref{compare}.  
\begin{equation} \label{Model 5}
\begin{split}
{\left[y_{\text{mean}}(\boldsymbol{s},t) \mid \mu_{\text{mean}}(\boldsymbol{s},t), \sigma_{\text{mean}}^2 \right] } &\sim \operatorname{Gaussian}\left(\mu_{\text{mean}}(\boldsymbol{s},t), \sigma_{\text{mean}}^2 \right)\\ 
\mbox{with}\qquad  \mu_{\text{mean}}(\boldsymbol{s},t) &  =   {\mathbf{x}^{S}(\boldsymbol{s},t)^\top\boldsymbol{\beta}^{S}}+{\mathbf{x}^{N\!S}(\boldsymbol{s},t)^\top \boldsymbol{\beta}_{\text{mean}}^{N\!S}}%\\&\quad
+{u^{S}(\boldsymbol{s},t)}; \\
{\left[y_{\text{max}}(\boldsymbol{s},t) \mid \mu_{\text{max}}(\boldsymbol{s},t), \sigma_{\text{max}}\right] } &\sim \operatorname{Gumbel}\left(\mu_{\text{max}}(\boldsymbol{s},t), \sigma_{\text{max}}\right) \\
 \mbox{with} \qquad\mu_{\text{max}}(\boldsymbol{s},t) &   =\ {\mathbf{x}^{S}(\boldsymbol{s},t)^\top\boldsymbol{\beta}^{S}}{\boldsymbol{\beta}_{1}^{\text{Gaussian-Gumbel}}}\\
&\quad+{\mathbf{x}^{N\!S}(\boldsymbol{s},t)^\top \boldsymbol{\beta}_{\text{max}}^{N\!S}} +{{\beta}_{2}^{\text{Gaussian-Gumbel}}}{u^{S}(\boldsymbol{s},t)}, \qquad
\end{split}
\end{equation}
where the terms $\boldsymbol{\beta}^{S}$ and $u^{S}(\boldsymbol{s},t)$ with superscript $S$ denote the sharing effects, and $\mathbf{x}^{S}(\boldsymbol{s},t)$ are corresponding variables (geographical and meteorological). The term $\boldsymbol{\beta}^{N\!S}$ with superscript $NS$ denotes the non-sharing effects with corresponding covariates $\mathbf{x}^{N\!S}(\boldsymbol{s},t)$ which includes the intercept, location variables and {human-effect variables} (longitude, latitude, {population density}). 
{For the selection of sharing and non-sharing variables, to avoid the potential issue of uncertainty of significant sharing effects (${\beta}^{S}=0$), we treat two significant predictors (altitude and precipitation) as sharing terms to investigate the potential similar and reverse effects by the ratios ($\boldsymbol{\beta}_1^{\text{Gaussian-Gumbel}}$) while considering all other variables, including three non-significant ones (temperature, vapour pressure, and population density), and location variables as non-sharing terms.}

Additionally, the spatio-temporal random effect is also treated as sharing effects with respect to the simplicity of computation. {The sharing effects $\boldsymbol{\beta}_1^{\text{Gaussian-Gumbel}}$ and $\boldsymbol{\beta}_2^{\text{Gaussian-Gumbel}}$ scale the common components of covariate vector $\boldsymbol{x}^S(\boldsymbol{s}, t)$ and random effect $u^S(\boldsymbol{s}, t)$, and control how much information is shared from the average predictor towards the annual max predictor, and determine the strength of interaction between the two processes. Precisely, it allows for capturing the positive or negative correlations. }

For further explanation, the sharing effects (including fixed effects and random effects) are the same in two sub-models ($\boldsymbol{\beta}^{S}$, $u^{S}(\boldsymbol{s},t)$). Meanwhile, they are linked by the sharing coefficients (scaling factors)  $\boldsymbol{\beta}_{1}^{\text{Gaussian-Gumbel}}$ and ${\beta}_{2}^{\text{Gaussian-Gumbel}}$. On the one hand, these sharing coefficients relax the strictly equal relation between the sharing effects in two sub-models. On the other hand, more importantly, their posterior distributions can also measure the similarities and differences in the effects of predictors in the sub-models. For instance, a significantly negative ${\beta}_{1}^{\text{Gaussian-Gumbel}}$ implies that the corresponding predictors oppositely influence the moderate and extreme air pollution cases.

\subsection{Priors definition}\label{Prior}
In a Bayesian context, in order to finalize the model, we need to define prior distributions for the remaining parameters in Gumbel ($\sigma_{\text{Gumbel}}$) and GEV distributions ($\sigma_{\text{GEV}}$, $\xi_{\text{GEV}}$), the regression coefficients ($\boldsymbol{\beta}$), the sharing coefficients in the joint model ($\boldsymbol{\beta}^{\text{Gaussian-Gumbel}}$), parameters in the Mat\'{e}rn covariance function ($\sigma_{M}, \rho_{M}$, $\nu_{M}$) and the parameters in AR(1) dynamic model ($a$, $\tau_{AR}$).

We use vague Gaussian priors for the tail parameter ($\xi_{\text{GEV}}$) in GEV distribution and the elements of coefficients ($\boldsymbol{\beta}$, $\boldsymbol{\beta}^{\text{Gaussian-Gumbel}}$). The smooth parameter $\nu_{M}$ is  treated here as a fixed value with $\nu_{M} = 1$, as in most spatial analyses. 
%The penalised complexity (PC) prior is a weakly informative prior distribution, designed to punish model complexity by placing an exponential prior on the distance from some base models.
The parameters $\sigma_{M}$ and $\rho_{M}$ in the Mat\'{e}rn function and autocorrelation parameter $a$ in AR(1) model {are defined by penalized complexity (PC) priors \citep{PC1} with knowledge from \cite{Geohealth} and \cite{PC2}.}  PC prior for the range parameter ($\rho_{M}$) is defined with $\operatorname{Prob}\left(\rho_{M}<10^4\right)=0.01$, which means the probability that the range is less than $10 \mathrm{km}$ is very small, and the PC prior for variance parameter as {$\operatorname{Prob} \left(\sigma_{M}>3\right)=0.01$,} 
indicating the probability for variance greater than $3$ is low. Similarly, we apply the {auto-correlation} ($a$) PC prior following the recommendation of  $\operatorname{Prob}\left(a>0\right)=0.9$. 

Note that INLA often uses the precision parameter ($\tau$) to replace the scale parameter ($\sigma$) by $\sigma  = {1}/{\sqrt{\tau}}$.  {The PC priors for all precision parameters are given by  $\operatorname{Prob}\left({1}/{\sqrt{\tau}}>3\right)=0.01$ for Gumbel and GEV likelihood and $\operatorname{Prob}\left({1}/{\sqrt{\tau_{AR}}}>5\right)=0.01$ for the AR model.}

\subsection{Model evaluation, diagnosis and cross-validation}\label{evaluation}
Traditional Bayesian model performance is evaluated by two popular criteria, the deviance information criterion (DIC) and the Watanabe-Akaike information criterion (WAIC). The deviance information criterion (DIC) proposed by \cite{DIC}, is a popular criterion for model choice similar to the Akaike information criterion (AIC).
$$
\text{DIC}=D( \widehat{\boldsymbol{\theta}})+2 p_D,
$$
where $D( \widehat{\boldsymbol{\theta}})$ is the deviance function with Bayes estimate $\widehat{\boldsymbol{\theta}}$, and $p_D$ is the effective number of parameters. The Watanabe-Akaike information criterion, also known as the widely applicable Bayesian information criterion, is similar to the DIC, but the effective number of parameters is computed in a different way \citep{WAIC}.

However, DIC may under-penalize complex models with many random effects. Alternatively, for prediction performance, INLA suggests applying the leave-one-out cross-validation criteria, conditional predictive ordinates \citep[CPO;][]{Petti} {with its summative version, the Logarithmic Score \cite[LS;][]{Gneiting2007}} and predictive integral transform \citep[PIT;][]{PIT2}, which facilitates the computation of the cross-validated log-score for model choice, and enables the calibration assessment of out-of-sample predictions, respectively.
$$
\begin{aligned}
{\text{LS}} &= - \sum_{i=1}^{n} \ln{\text{CPO}_i} \quad \text{with\ } \mathrm{CPO}_i =\pi\left(y_i^{\text {obs }} \mid y_{-i}\right),\\
\text{PIT}_i &=\operatorname{Prob}\left(Y_i \leq y_i^{\text {obs }} \mid y_{-i}\right),% .
\end{aligned}
$$
where $y_i^{\text {obs }}$ denotes the $i$-th observation and $y_{-i}$ denotes the observations $y$ with the $i$-th component omitted. {A model with a small value of LS and a standard uniform distributed PIT is preferable \citep{GR2020}.}

{Note that} numerical problems may occur when CPO and PIT values are computed with the complicated Gaussian process with INLA algorithm \citep{CPOINLA}. {Hence}, we take a few other criteria introduced in Bayesian literature into account: The coverage probability of 95\% CI is computed as the proportion of the validation observations that the observed value lies between the 2.5\% quantile to the 97.5\% quantile of the predicted value (posterior distribution). The correlation coefficient denotes the correlation between observed values and predicted values in the validation set. The root mean square error (RMSE) is defined as the square root of the second sample moment of the differences between predicted values and observed values.

As we separate the original five-year dataset (2017--2021) into training (2017--2020) and validation (2021) sets, we decide to apply DIC, WAIC, {LS}, PIT and RMSE to compare the model fitness on the training set and use the coverage probability, correlation coefficient, and RMSE for predictive ability evaluation.

\section{Results}\label{RandD}

In this section, we firstly compare the performance of the Bayesian generalised extreme models and select the best {one} with the outstanding predictive ability (Section \ref{compare}), then summarize the corresponding posterior estimates for both fixed and random effects (Section \ref{EVM output}). {The analysis and interpretation of the joint model are included in Section \ref{Joint output} and followed by the excursion functions generation with severe air pollution risk ranking (Section \ref{Prediction}).}
Note that since the explanatory variables are measured in different scales, to avoid numerical problems, each predictor is standardized to have mean zero and unit standard deviation.

\subsection{Model comparison}\label{compare}
\begin{figure} [H]
	\centering
	\subfloat[\label{fig3:a}]{
		\includegraphics[scale=0.35]{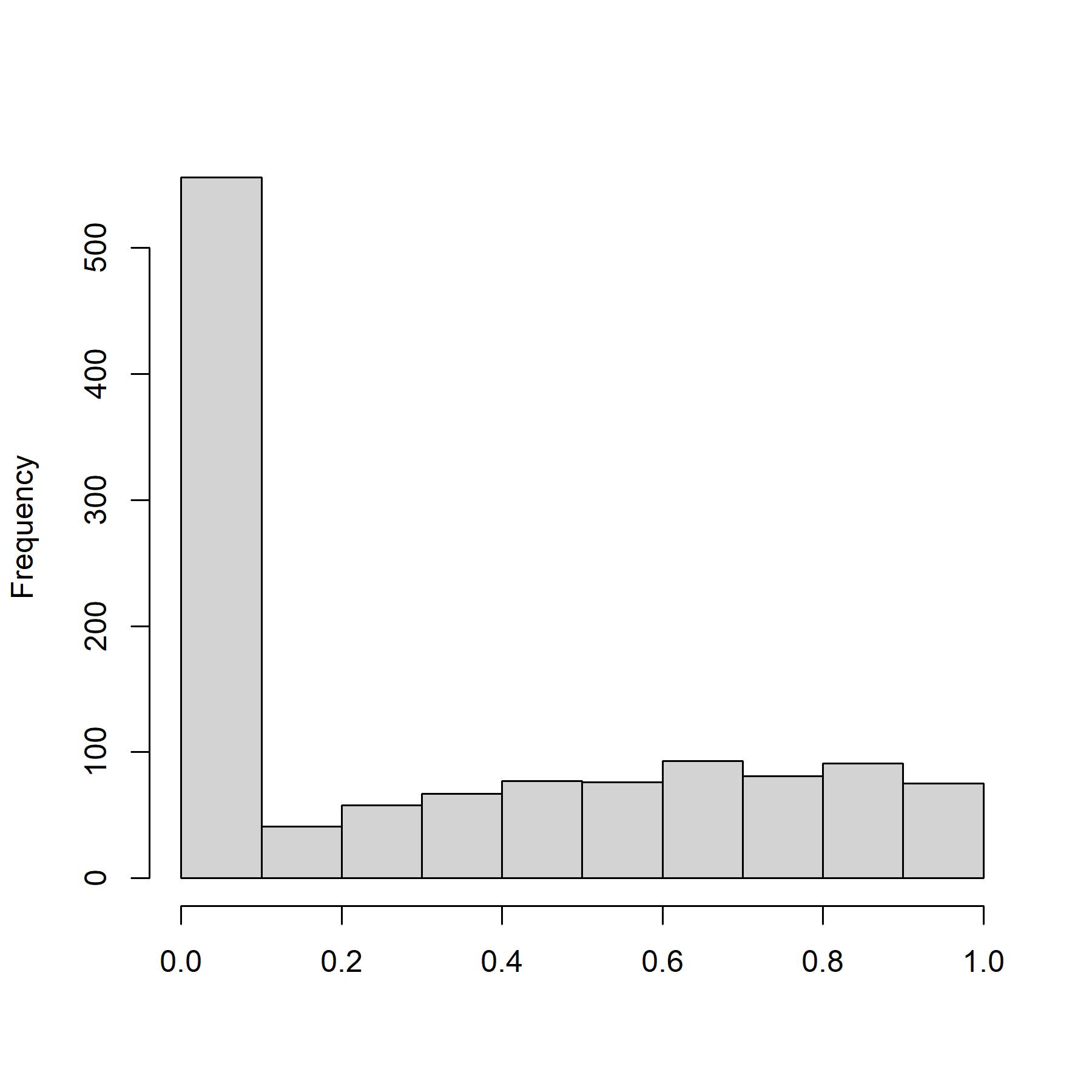}}
	\subfloat[\label{fig3:b}]{
		\includegraphics[scale=0.35]{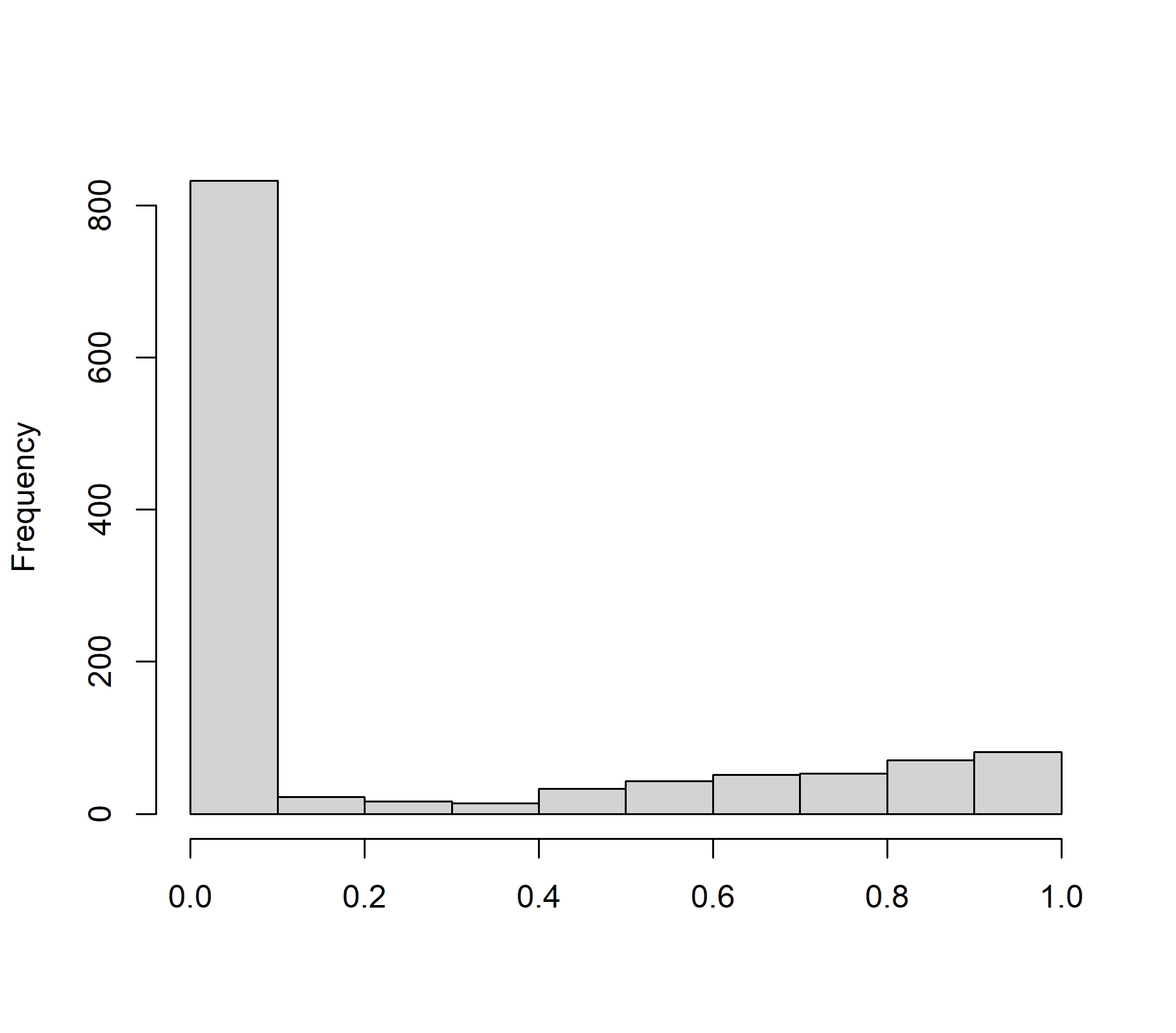}}
	\\
	\subfloat[\label{fig3:c}]{
		\includegraphics[scale=0.35]{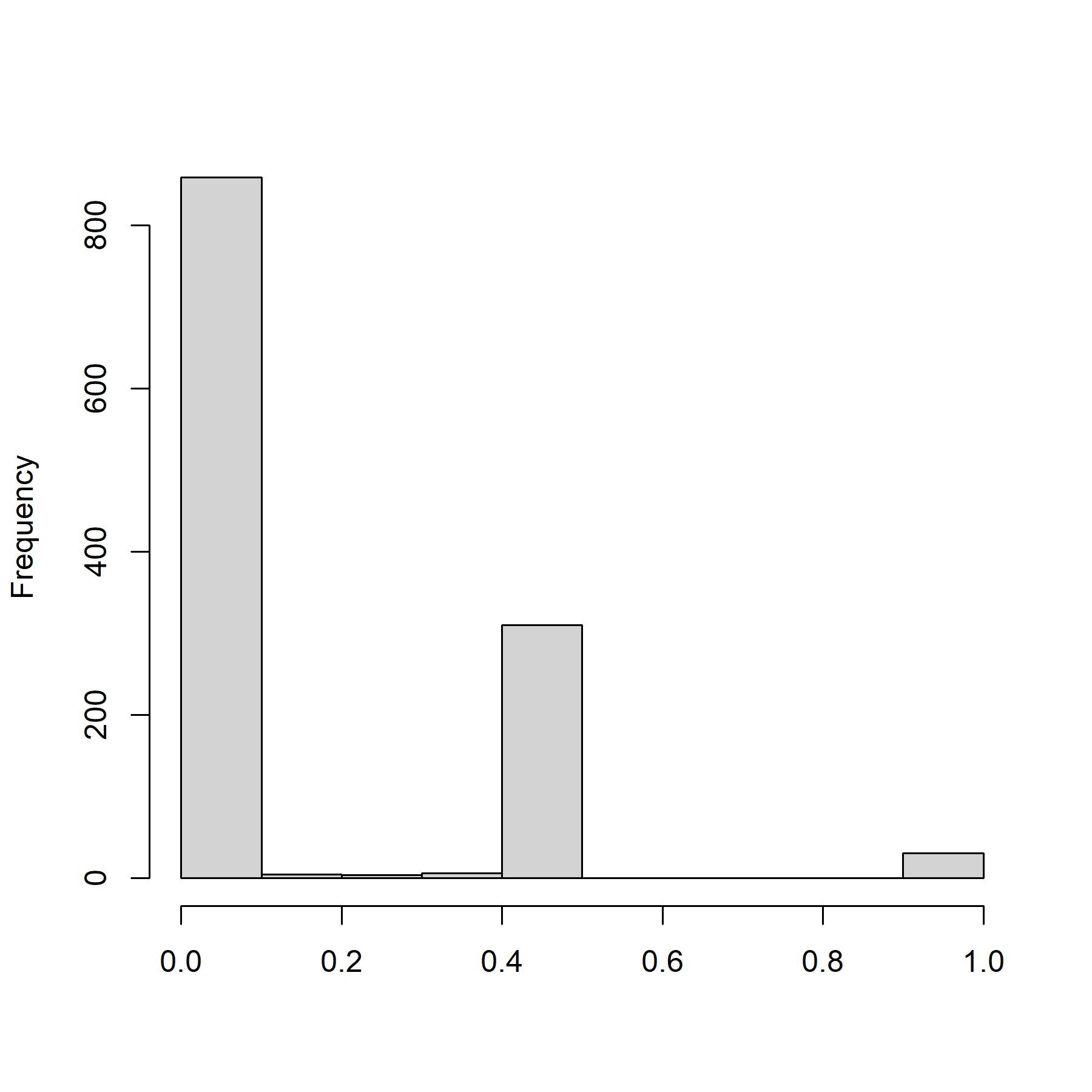} }
	\subfloat[\label{fig3:d}]{
		\includegraphics[scale=0.35]{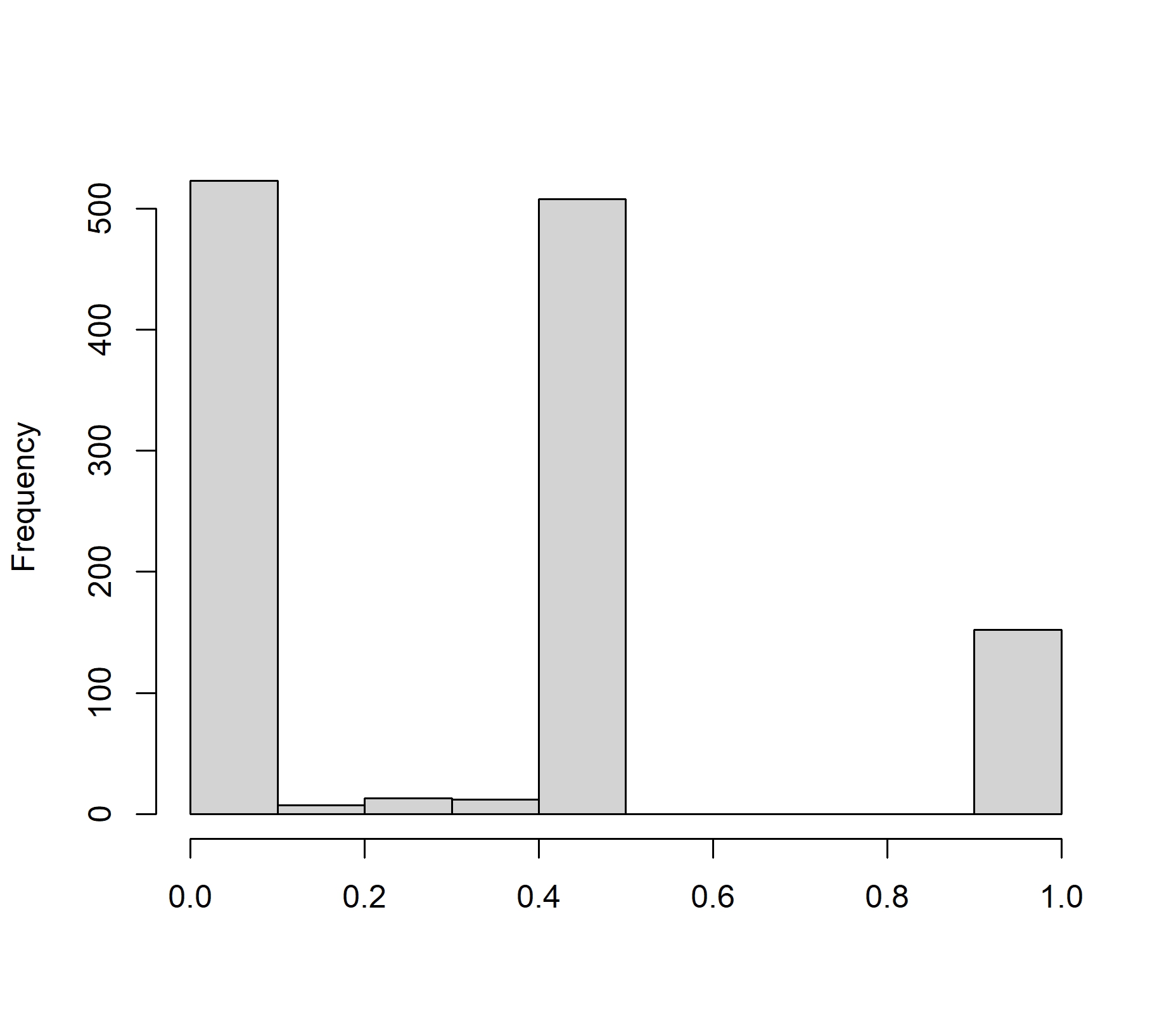}}
	\caption{Predictive integral transform plots for Models 1 $\sim4$ {specified in Eqs.\eqref{Model 1}, \eqref{Model 2}--\eqref{Model 4}} 
 in (a)$\sim$(d) {based on training dataset}. A uniform distribution pattern is preferable.} 
	\label{PIT} 
\end{figure}

In order to examine the fitting and prediction ability of the extreme value models, we apply some evaluation criteria (DIC, WAIC, {LS}, PIT and RMSE) on the training set (Table \ref{Criteria}), and use other criteria (coverage probability, correlation and RMSE) on the validation set (Table \ref{val}). {{We see from} Table \ref{Criteria} {with model} %of 
performance on the training set, Models 2 and 4 outperform in DIC (36.27 and 73.43, respectively), WAIC (193.67 and 223.06, respectively) and RMSE (0.24 and 0.18, respectively). However, Model 1 is preferable in the leave-one-out cross-validation criteria with {LS (8019.07)} and PIT plots (Figure \ref{PIT}). %In Table \ref{val} about the validation set, 
{Table \ref{val} shows that} all four models are comparable in the coverage probability, correlation and RMSE.}

\begin{figure*} [htbp!]
	\centering
	\subfloat[\label{fig4:a}]{
		\includegraphics[scale=0.3]{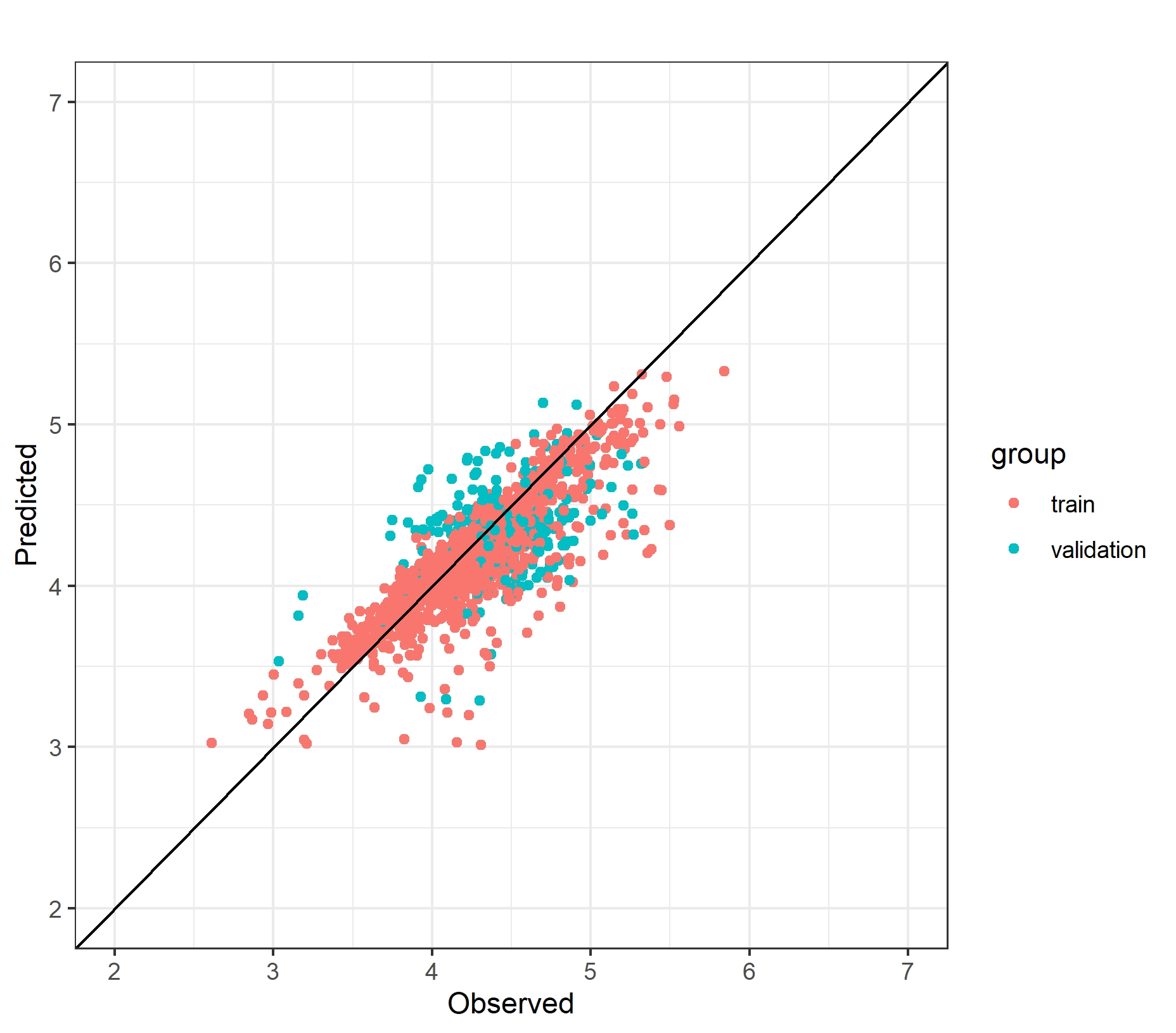}}
	\subfloat[\label{fig4:b}]{
		\includegraphics[scale=0.3]{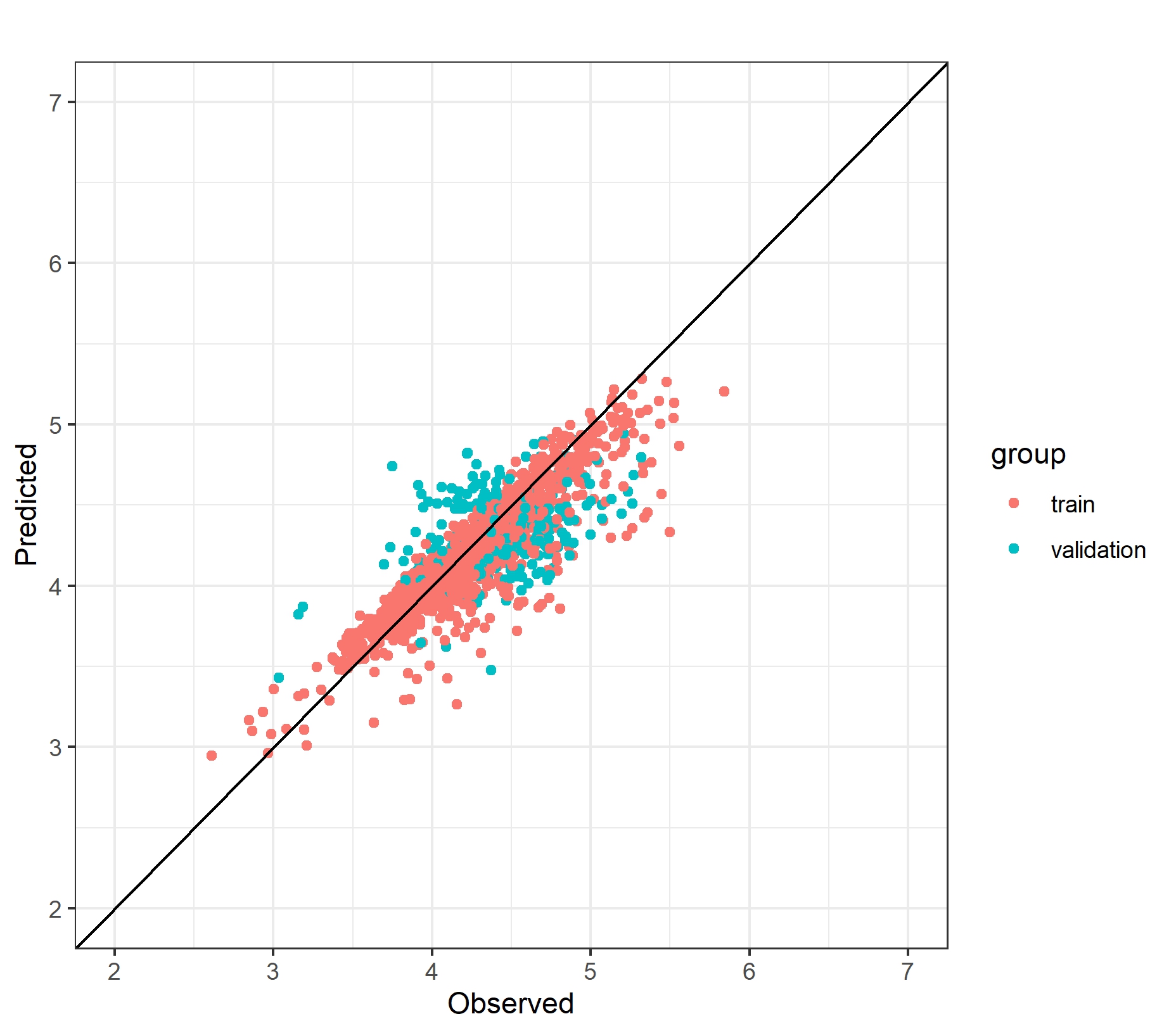}}
	\\
	\subfloat[\label{fig4:c}]{
		\includegraphics[scale=0.3]{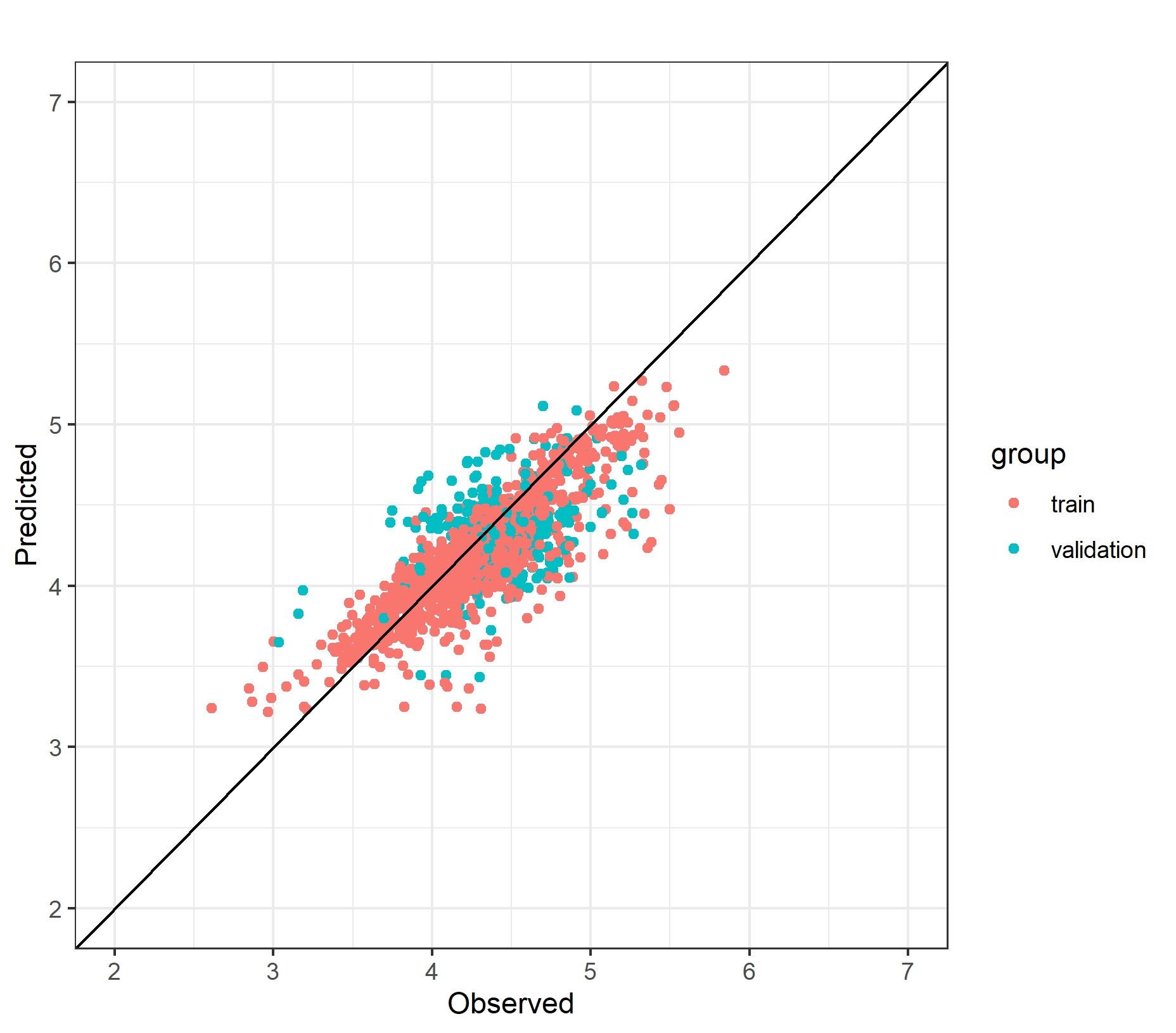} }
	\subfloat[\label{fig4:d}]{
		\includegraphics[scale=0.3]{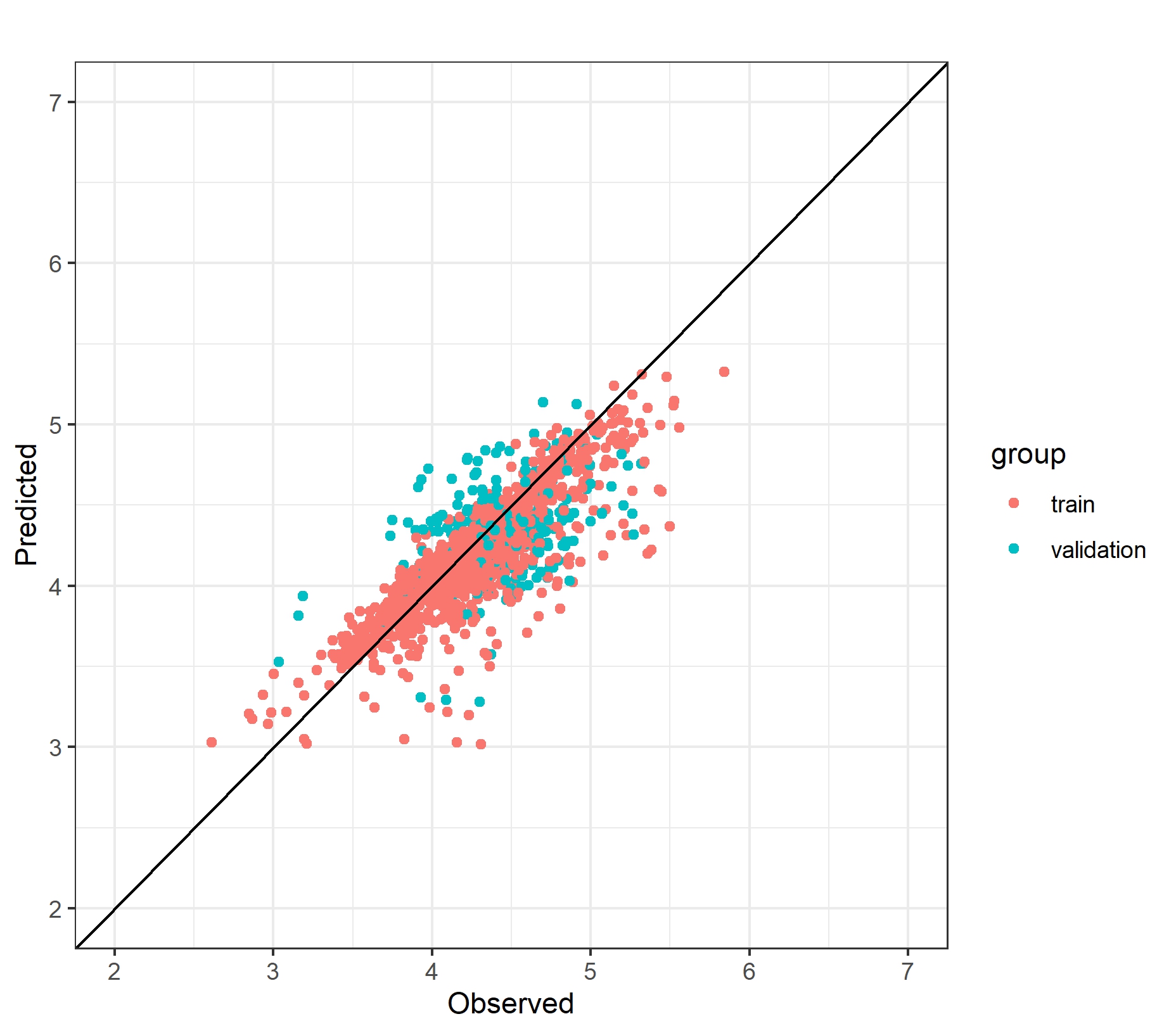}}
	\caption{{Visualisation} of training (\textcolor{red}{red}) and validation (\textcolor{cyan}{cyan}) performance {in  (a)$\sim$(d) for Models 1$\sim$4 specified in Eqs.\eqref{Model 1}, \eqref{Model 2}--\eqref{Model 4} subsequently.} The scatters distributed along the line with the intercept 0 and the slope denote the better model.}
	\label{TV plot} 
\end{figure*}

{Figure \ref{TV plot} shows the simultaneous visualization of model performance on both training and validation sets, where {the scatters' distribution} along the line with intercept 0 and slope 1 indicates the estimated (predicted) values are best suited to the observations. All four models exhibit a strong fit for the trend.}

{To summarise, no model excels in all criteria. Considering the similar performance of all four models on the validation set, we choose the parsimonious model, Model 1 (the Gumbel model with fixed effects and spatio-temporal random effects defined in Eq.\eqref{Model 1}), as the best model and bring its structure into further analysis.}

\begin{table}[H]
\centering
\small    
\caption{Performance evaluation criteria (DIC, WAIC, {LS} and RMSE) {of Models 1$\sim4$ specified in Eqs.\eqref{Model 1}, \eqref{Model 2}--\eqref{Model 4}} on the training set (all criteria prefer lower values).}
  \begin{tabular*}{\textwidth}{@{\extracolsep{\fill}}crrrc}
    \toprule  Model & DIC & WAIC & LS & RMSE \\
    \midrule Model 1  & 229.78 & 356.86 & 8019.07 & 0.24 \\
           %{Model 2}  & {654.45} & {829.47} & {19969.64} & {0.24} \\
            { Model 2} & {36.27} & {193.67} & {13140.13} & {0.20} \\
           Model 3 & 277.48 & 317.67 & 12128.54 & 0.23\\
           %Model 4 & {654.86} & {792.23} & {18205.71} & {0.29} \\
         {Model 4} & {73.43} & {123.06} & {14314.97} & {0.18} \\
 \bottomrule
    \end{tabular*}
    \label{Criteria}
\end{table}

\begin{table}[htbp!]
\centering
\small    
\caption{Performance evaluation criteria (coverage probability, correlation and RMSE) {of Models 1$\sim4$ specified in Eqs.\eqref{Model 1}, \eqref{Model 2}--\eqref{Model 4}} on the validation set. High coverage probability, high correlation, and low RMSE indicate the preferable model.}
\begin{tabular*}{\textwidth}{@{\extracolsep{\fill}}cccc}
    \toprule  Model & Coverage Probability & Correlation  & RMSE \\
    \midrule Model 1  & 83.95\% & 40.95\% & 0.36  \\
           %Model 2  &  {74.11\%} & {41.20\%} & {0.43} \\
            { Model 2} & {84.60\%} & {39.19\%} & {0.35}  \\
           Model 3 & 83.90\% & 37.21\% & 0.37 \\
          % Model 4 & {78.43\%} & {43.84\%} & {0.40} \\
         { Model 4} & {86.97\%} & {37.33\%} & {0.36}  \\
 \bottomrule
    \end{tabular*}
    \label{val}
\end{table}

\subsection{Summary of Model 1}\label{EVM output}

The summary statistics for fixed effects are shown in Table \ref{M1summary}. {We see that} %For explanatory variables, 
altitude and population density are significantly negatively associated with annual maxima $\mathrm{PM}_{10}$ concentrations, whereas the effects of other covariates are not statistically significant. High temperature, low precipitation and low vapour pressure are likely to associate with extreme $\mathrm{PM}_{10}$ concentrations, {consistent with \cite{KALISA2018} and \cite{LI2014}.} 

The negative association between population density and annual maxima $\mathrm{PM}_{10}$ concentrations is different from the positive correlation found in Table \ref{corr} and the positive association demonstrated by \cite{BORCK2021103596}. Together with the facts of relatively low correlation for population density (Table \ref{corr}) and the phenomena that severe $\mathrm{PM}_{10}$ pollution in 2017, 2020 and 2021 (Figure \ref{pm10}) is usually associated with high temperature and low precipitation, our results probably imply {that} the generation and spread of extreme $\mathrm{PM}_{10}$ concentrations depend heavily on the climate conditions. {This evidence coincides with the findings of the overwhelming role of adverse meteorology in severe air pollution events \citep{WANG2020,MORAWSKA2021}.}

\begin{figure*} [ht]
	\centering
	\subfloat[\label{fig5:a}]{
		\includegraphics[scale=0.3]{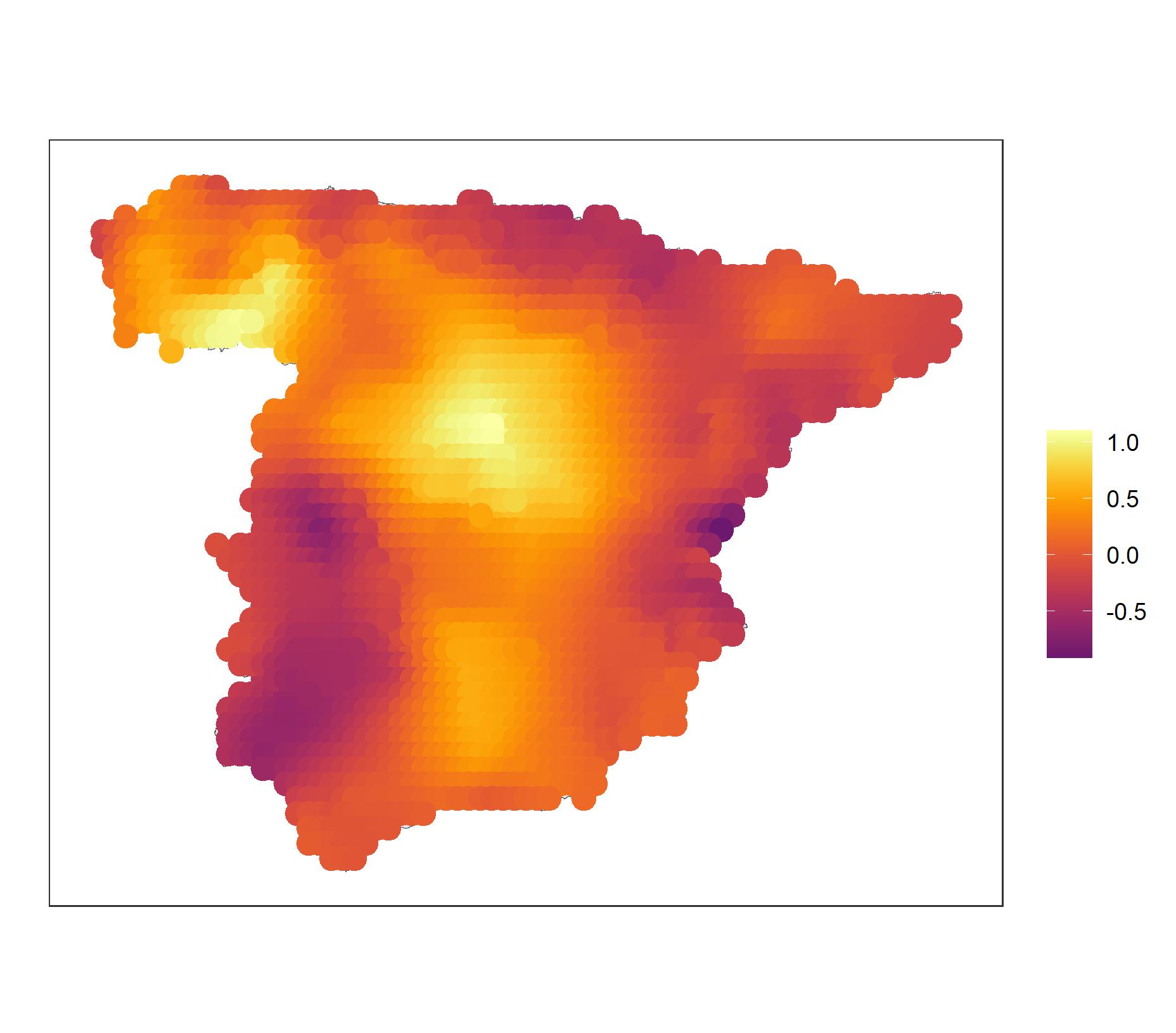}}
	\subfloat[\label{fig5:b}]{
		\includegraphics[scale=0.3]{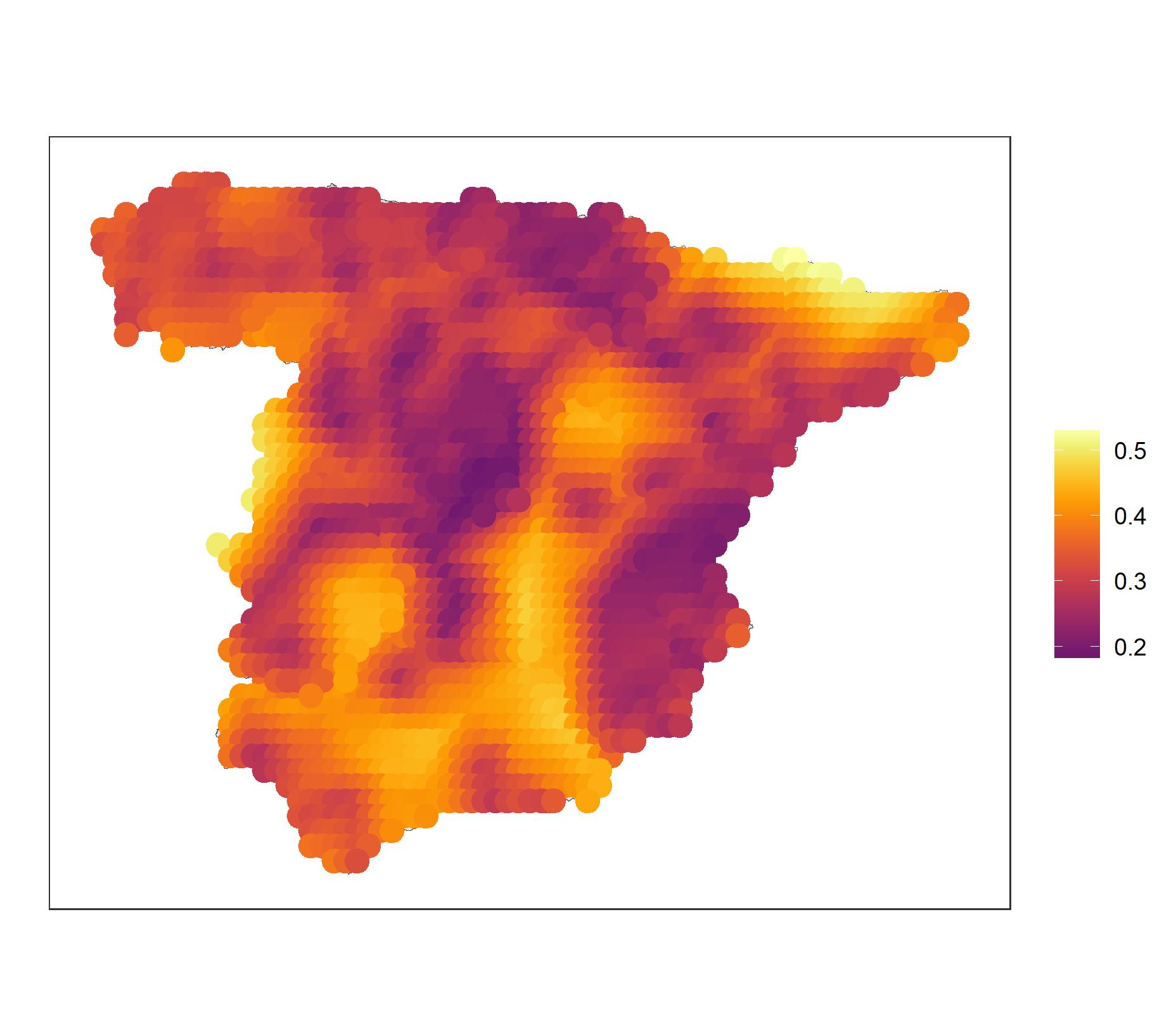}}
	\caption{Heat map of the spatial random effect in 2017 with (a) mean and (b) standard deviation.}
	\label{Spatial pattern} 
\end{figure*}

The spatial and temporal dependence is accessible by spatial heat plot (Figure \ref{Spatial pattern}) and posterior estimates of autocorrelation coefficient ($a$ in Table \ref{M1hyper}). Spatially, similar values of mean random effects occur in groups, especially, a large cluster of high values happens in the centre of the mainland. Temporally, the estimated mean correlation coefficient ($a$) is 0.80 with  95\% credible interval (0.74, 0.85),  
providing evidence of relatively strong dependence between two consecutive years.

\begin{table}[ht]
    \centering
\small    
\caption{Posterior estimates (mean, standard deviation and quantiles) of the coefficients of the covariates involved in Model 1 {specified in Eq.\eqref{Model 1}}.}
\setlength{\tabcolsep}{0.2mm}
\begin{tabular*}{\textwidth}{@{\extracolsep{\fill}}lrrrrrr}
    \toprule  Covariate  & Mean & Stdev & 0.025 quantile & 0.5 quantile & 0.975 quantile \\
    \midrule Intercept  & $4.20$ & $0.12$ & $3.96$ & $4.20$ & 4.43 \\
           Longitude  & $-0.08$ & 0.13 & $-0.33$ & $-0.08$ & 0.17 \\
           Latitude & $-0.06$ & 0.13 & $-0.31$ & $-0.06$ & 0.19\\
           Altitude & $-0.08$ & 0.03 & $-0.13$ & $-0.08$ & $-0.03$\\
           Temperature & 0.13 & 0.10 &  $-0.07$ &  0.13 & 0.32\\
           Precipitation & $-0.03$ & 0.06 & $-0.15$ & $-0.03$ & 0.09\\
           Vapour Pressure & $-0.09$ & 0.09 & $-0.27$ & $-0.09$ & 0.09\\
           Population Density & $-0.07$ & 0.03 & $-0.13$ & $-0.07$ & $-0.01$\\
 \bottomrule
    \end{tabular*}
    \label{M1summary}
\end{table}

\begin{table}[ht]
    \centering
\small    
\caption{Posterior estimates of mean, standard deviation and quantiles  of the parameters {in Model 1 specified in Eq.\eqref{Model 1}.} The precision ($\tau_{G}$) for Gumbel distribution, the range parameter ($\rho_{M}$) and the standard deviation ($\sigma_{M}$) introduced in Mat\'{e}rn covariance function  and the {autocorrelation} coefficient ($a$). }
\setlength{\tabcolsep}{0.15mm}
\begin{tabular*}{\linewidth}{@{\extracolsep{\fill}}lrrrrr}
    \toprule  Parameter  & Mean & Stdev & 0.025 quantile & 0.5 quantile & 0.975 quantile \\
    \midrule Precision ($\tau_{G}$)  & 25.62 & 1.64 & 22.52 & 25.57 & 28.99 \\
           Range ($\rho_{M}$, unit $10^5$)  & $2.62 %\times 10^5
           $ & %$0.374\times 10^4$ 
           0.37
            & $1.96 %\times 10^5
           $ & $2.60% \times 10^5
           $ & $3.43% \times 10^5
           $ \\
           Stdev ($\sigma_{M}$) & 0.58 & 0.05 & 0.48 & 0.57 & 0.69\\
           Autocorrelation ($a$) & 0.80 & 0.03 & 0.74 & 0.80 & 0.85 \\
 \bottomrule
    \end{tabular*}
    \label{M1hyper}
\end{table}

\subsection{Results of the joint model with sharing effects}\label{Joint output}

Combining the best extreme value model (Model 1) with the Gaussian model, we establish the joint model in Section \ref{joint} with sharing effects to estimate both maxima and mean concentration levels simultaneously. {We keep the two significant effects of predictors (altitude and precipitation) as sharing effects, and all other effects (temperature, vapour pressure, population density and location) are treated as non-sharing ones.} {Table \ref{sharing} summarizes the posterior estimates of these effects from the joint model.} Precipitation shows significant negative associations with both annual mean and maxima concentrations, {which is consistent with the notion that $\mathrm{PM}_{10}$ concentrations are generally low in wet regions \citep{LI2014}}. {In contrast, we find that altitude is negatively connected with the mean but positively connected with the maxima.}  

\begin{figure*} [ht]
\centering
	\subfloat[\label{fig8:a}]{
		\includegraphics[scale=0.45]{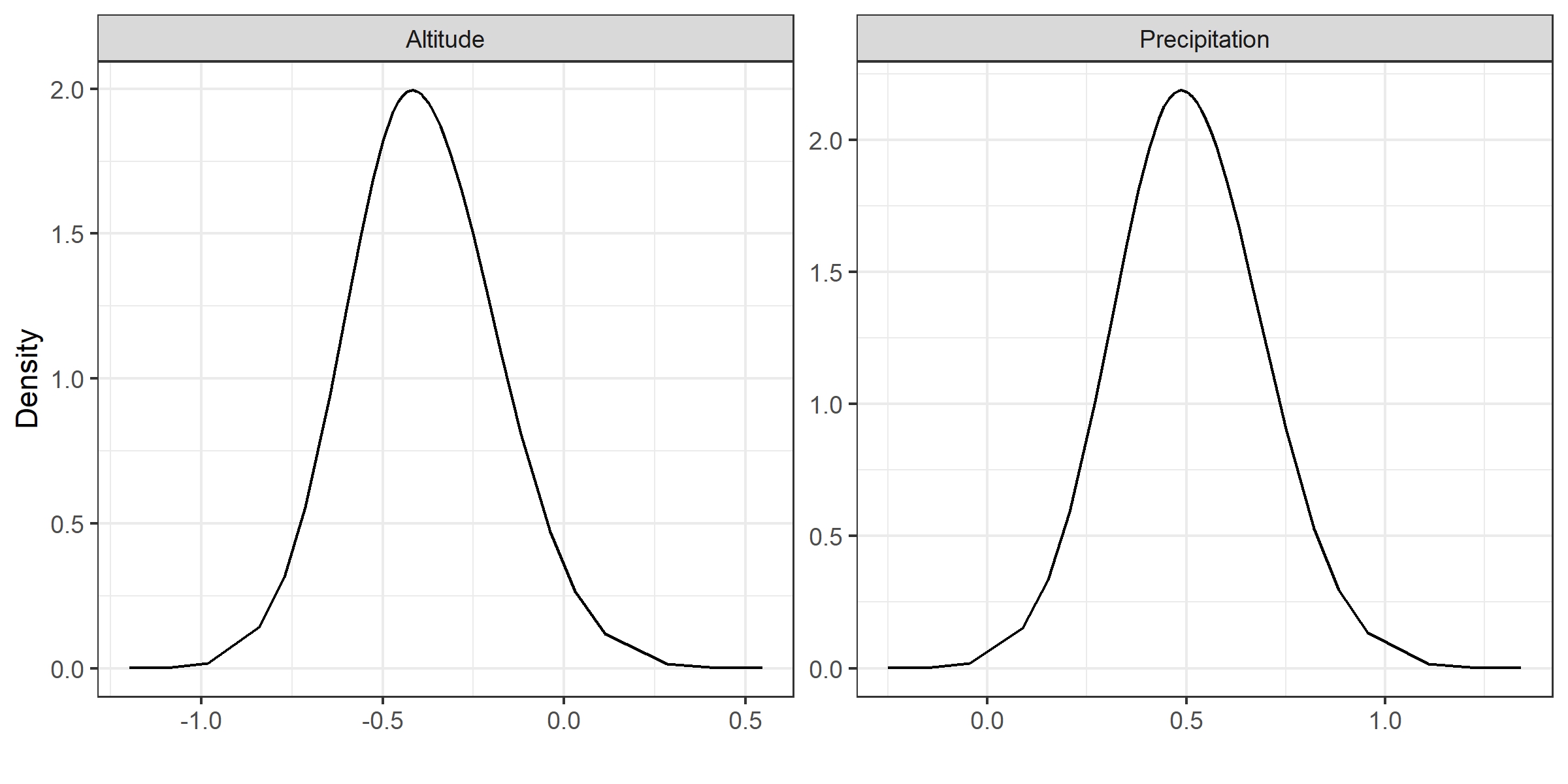}}\\
	\subfloat[\label{fig8:b}]{
		\includegraphics[scale=0.5]{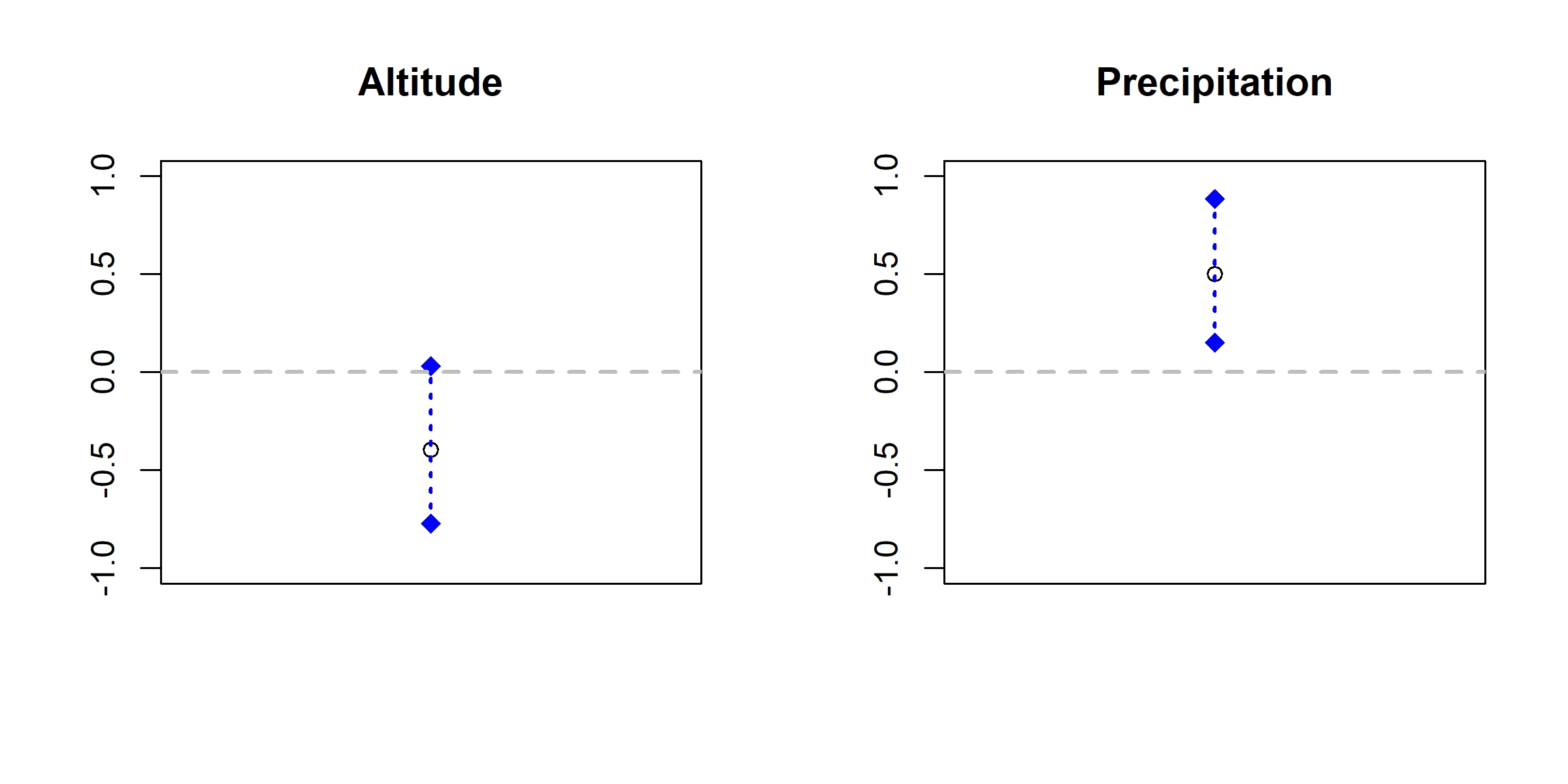}}
\caption{(a) The posterior distributions plot and (b) the quantiles plot of sharing coefficients. The three nodes in the quantiles plot indicate 0.025, 0.5 and 0.975 quantiles of the posterior estimates.}
	\label{Beta plot} 
\end{figure*}

{This is also supported by the detailed sharing coefficients analysed below. We see from Figure \ref{Beta plot} that the posterior distribution of the sharing coefficients of precipitation (see detailed definition of $\boldsymbol{\beta}_{1}^{\text{Gaussian-Gumbel}}$  in Section \ref{joint}) almost lies between 0 and 1, showing that the influences of precipitation are similar on extreme and moderate {air} pollution. The coefficient for altitude is negative, suggesting certain evidence that altitude can impact reversely in different scaled pollution.}

{Furthermore, the effect of temperature is not significant with respect to 95\% credible interval for mean but becomes significant for maxima. This result implies that high temperatures weakly impact the general air quality, while playing an essential role in the dispersion of extremely poor pollution. Nevertheless, considering the possibility of unmeasured confounders potentially distorting the coefficients, the strength of the aforementioned inference might be compromised.
}

\begin{figure*} [ht]
	\centering
	\subfloat[\label{fig9:a}]{
		\includegraphics[scale=0.25]{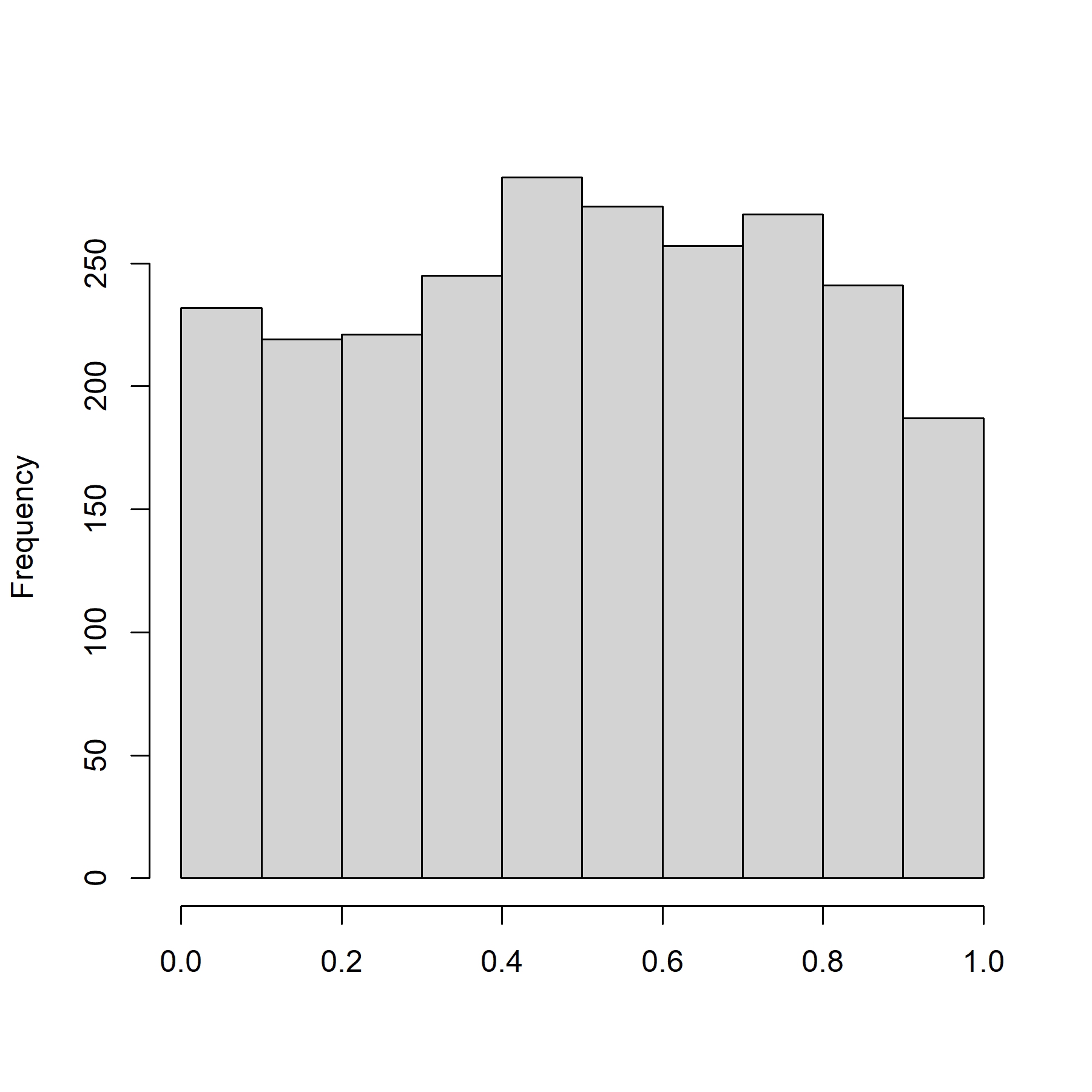}}
	\subfloat[\label{fig9:b}]{
		\includegraphics[scale=0.25]{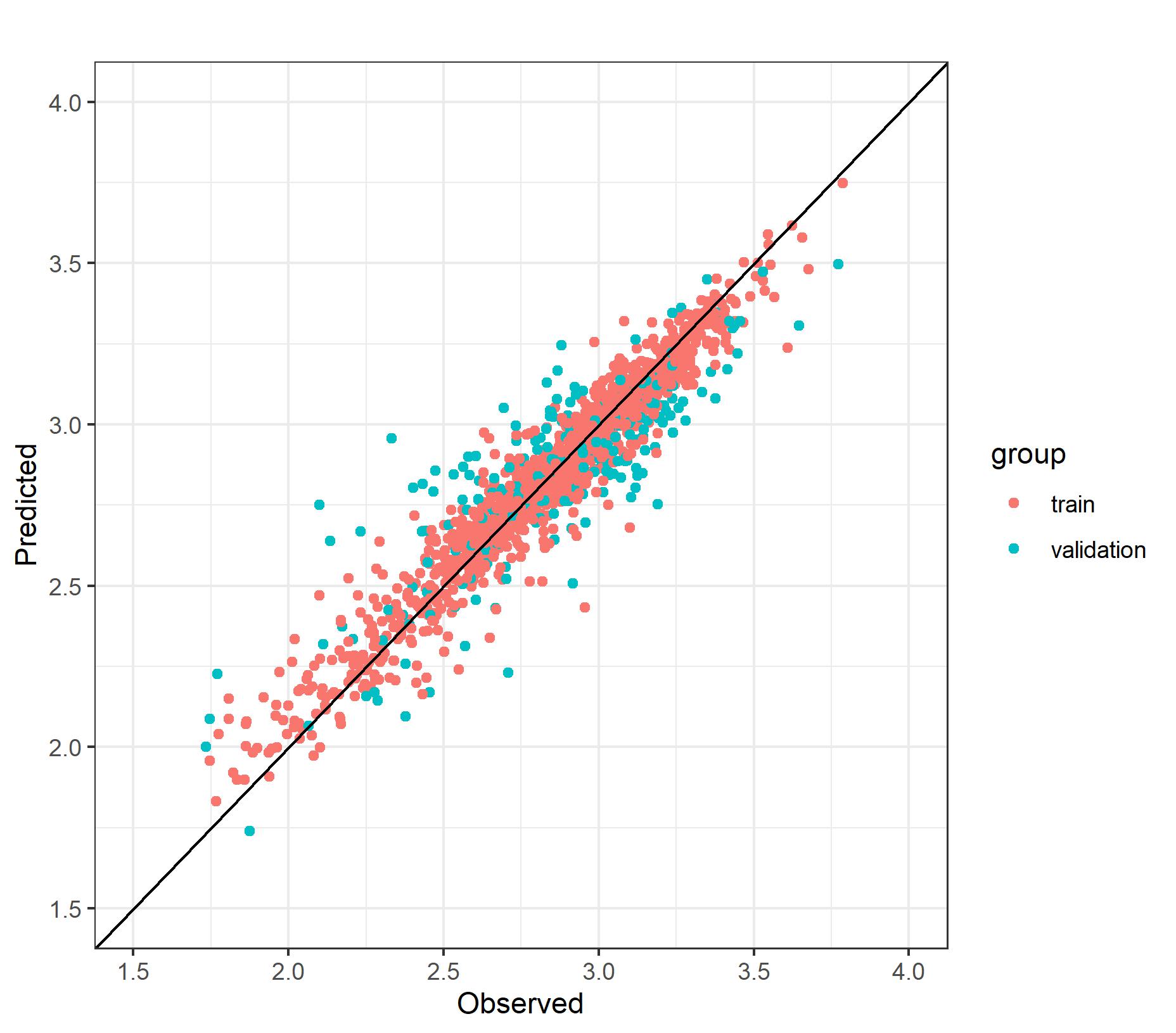}}
        \subfloat[\label{fig9:c}]{
		\includegraphics[scale=0.25]{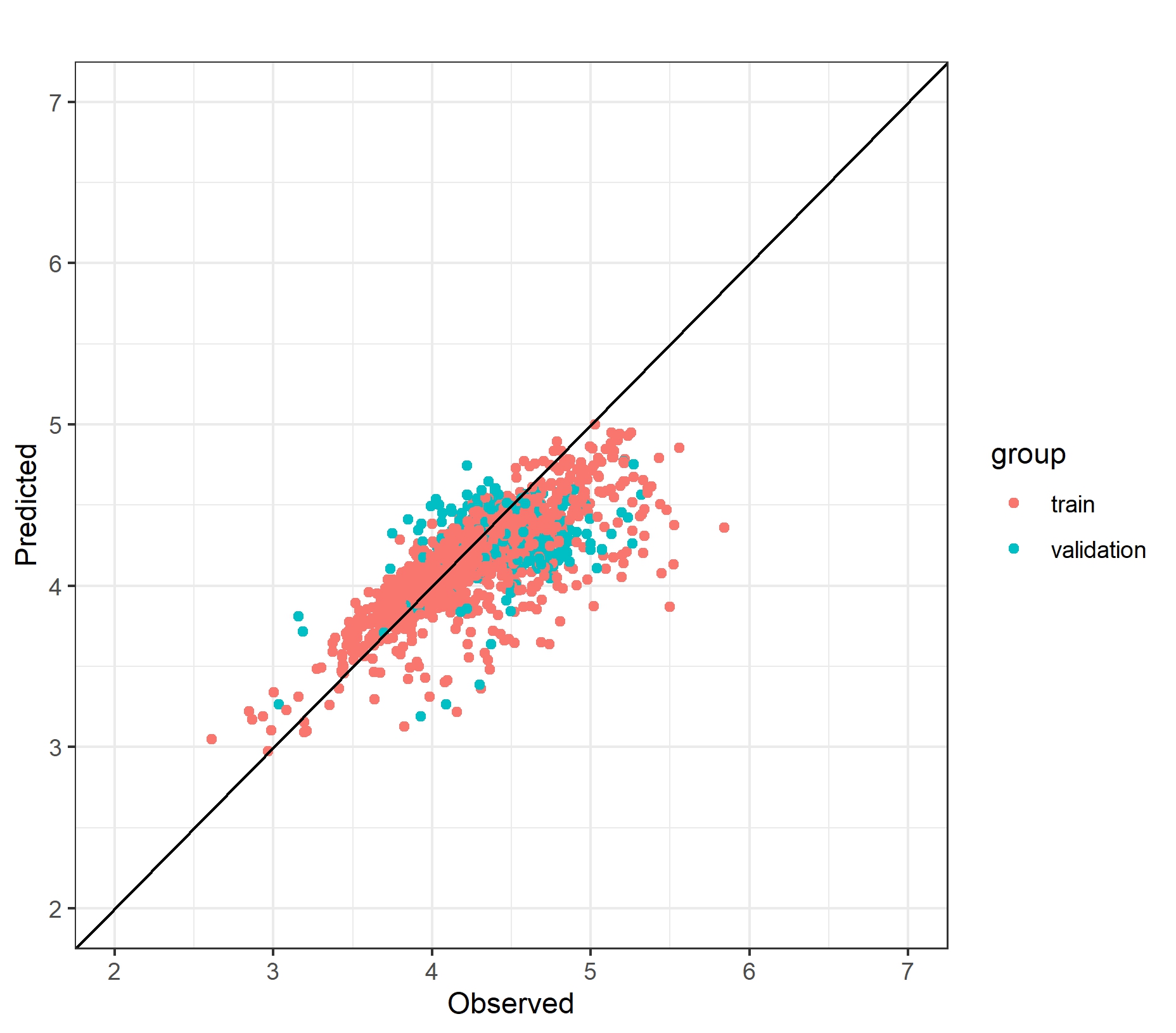}}
\caption{(a) Histogram of PIT for joint model {defined in Eq.\eqref{Model 5}} on training set, (b) the mean sub-model and (c) the max sub-model performance in both training and validation sets.}
	\label{valid plot} 
\end{figure*}

%The {posterior} distribution of the precipitation sharing coefficient 
{In terms of model performance evaluation, the joint model demonstrated promising results in the validation set. In particular,  the Gumbel sub-model (max sub-model) exhibited improvements in the coverage probability ($88.63\%$) and correlation ($46.71\%$), compared with the univariate Gumbel model (Model 1), as shown in Tables \ref{val} and  \ref{submodel}. The insights of the joint model performance are depicted in Figure \ref{valid plot}, where both
the performance of the Gumbel model (maxima model) and the Gaussian model (mean model) are satisfied. Because of this, despite the less parsimonious nature of the joint model, which may introduce more volatility in the prediction of the Gumbel sub-model, the overall satisfying evaluation results bolster the credibility of the joint model's findings. With this confidence, we proceed to utilize the joint model for subsequent predictions using excursion functions.}

%As for model performance evaluation, the joint model performed well in the training set. 

%{However, this is because the assessment criteria simultaneously consider both sub-models (Gaussian and Gumbel), and the well-fitted Gaussian model for the annual mean PM$_{10}$ can positively impact the extreme Gumbel model through sharing effects.} \highlighting{Please add some comments on the improved performance of the Gumbel-sub model compared with the univarariate Gumbel in Model 1, e.g., the coverage probability (88.63\%>83.95\%, 46.71\%>40.96\% correlation and comparable RMSE, see e.g., Table \ref{val} and \ref{submodel}}.

%More specifically, Table \ref{submodel} and Figure \ref{valid plot} provide insight into the validation performance of the two sub-models, where the performance of the Gumbel model (maxima model) is relatively good but less comparable to the Gaussian model (mean model). 

%{In particular, the prediction ability is improved with increased correlation as 56.10\% and the PIT plot tends to be uniform distributed in the joint model, see details in Table \ref{submodel} and Figure \ref{valid plot}(a).}

\begin{table}[ht]
 \centering
\caption{Posterior estimates (mean and 95\% credible interval) of the sharing effects and non-sharing effects specified in Eq.\eqref{Model 5}.}
\setlength{\tabcolsep}{0.5mm}
\resizebox{\linewidth}{!}{  
\begin{tabular}{*{5}{lrcrc}}
\toprule
  \multirow{2}*{Covariate} & \multicolumn{2}{c}{Annual mean model ($\boldsymbol{\beta}^{S}$)} & \multicolumn{2}{c}{Annual maxima model ($\boldsymbol{\beta}^{S}\boldsymbol{\beta}_{1}^{\text{Gaussian-Gumbel}}$)}  \\
  \cmidrule(rr){2-3}\cmidrule(lr){4-5}
  & Mean & 95\% CI & Mean &  95\% CI  \\
  \midrule
  {Altitude} & $-0.113$ & $ (-0.136, -0.090)$ & $0.047$ & $(0.037, 0.058)$  \\
  {Precipitation} & $-0.080$ & $(-0.107, -0.054)$ & $-0.039 $ & $ (-0.054, -0.026)$  \\
  \hline
   \multirow{2}*{Covariate} & \multicolumn{2}{c}{Annual mean model ($\boldsymbol{\beta}_{\text{mean}}^{N\!S}$)} & \multicolumn{2}{c}{Annual maxima model ($\boldsymbol{\beta}_{\text{max}}^{N\!S}$)}  \\
  \cmidrule(rr){2-3}\cmidrule(lr){4-5}
  & Mean & 95\% CI & Mean &  95\% CI  \\
  \midrule
  {Intercept} & $2.780$ &$ (2.726, 2.835)$ & $4.015$ &$ (3.966, 4.065)$ \\
  {Longitude} & $-0.007$ &$ (-0.059,  0.045)$ & $-0.066$ &$ (-0.114, -0.018)$ \\
  {Latitude} & $ -0.046$ & $ (-0.111, 0.020)$ & $ -0.015$ &$ (-0.077, 0.048)$ \\  
  {Temperature} & $0.059$ & $ (-0.029, 0.146)$ & $0.154$ & $ (0.066, 0.242)$  \\
   {Vapour Pressure} & $-0.053$ & $ (-0.134, 0.027)$ & $-0.071$ & $ (-0.148, 0.007)$ \\
  {Population Density} & $-0.047$& $ (-0.110, 0.016)$ & $-0.034$ &$ (-0.090, 0.021)$ \\
  \bottomrule
\end{tabular}}
\label{sharing}
\end{table}

\begin{table}[ht]
    \centering
\small    
\caption{Sub-models performance evaluation (coverage probability, correlation and RMSE) in the validation set.}
  \begin{tabular*}{\textwidth}{@{\extracolsep{\fill}}lrrrr}
    \toprule  Sub-model & Coverage Probability & Correlation & RMSE \\
    \midrule Gaussian Model (Mean)  & 78.82\% & 84.01\% & 0.19  \\
     Gumbel Model (Maxima)  & {88.63\%} &  {46.71\%} & {0.37} \\
 \bottomrule
    \end{tabular*}
    \label{submodel}
\end{table}

\subsection{Annual maxima prediction and excursion functions}\label{Prediction}

Hot-spot region identifications and predictive concentration level plots are important tools in air pollution studies, because they intuitively imply the air quality in specific regions and are easily interpreted to environmental agencies and the general public. \cite{excursion} proposed the positive ($\mathrm{E}_{u, \alpha}^{+}(X)$) and negative excursion sets ($\mathrm{E}_{u, \alpha}^{-}(X)$) that determine the largest set that simultaneously exceeds or below the risk level ($u$) with a {small error}
%minimum
probability ($\alpha$), employing a parametric family and sequential importance sampling method for estimating joint probabilities. To visualize the excursion sets simultaneously, we apply the positive and negative excursion functions, $F_u^{+}(s)=1-\inf \left\{\alpha \mid s \in \mathrm{E}_{u, \alpha}^{+}\right\}$ and $F_u^{-}(s)=1-\inf \left\{\alpha \mid s \in \mathrm{E}_{u, \alpha}^{-}\right\}$. To explain, the term $\inf \left\{\alpha \mid s_0 \in \mathrm{E}_{u, \alpha}^{+}\right\}$ denotes the "smallest" $\alpha$ required that the location ($s_0$) can be included into the positive excursion set $\mathrm{E}_{u, \alpha}^{+}$ at the first time, while the higher $1-\inf \left\{\alpha \mid s_0 \in \mathrm{E}_{u, \alpha}^{+}\right\}$ reported by positive excursion function generally indicates higher probabilities for the location ($s_0$) to exceed the risk threshold $u$ {simultaneously}.

%Please check again the definition/idea of excursion function since here does not convince me of the meaning of $p$-value. 

In our case, to discover the areas that are most likely and most unlikely to suffer from severe $\mathrm{PM}_{10}$ pollution {simultaneously}, we utilize predicted $\mathrm{PM}_{10}$ concentration levels from {the max sub-model of the joint model} to generate both positive and negative excursion functions with the thresholds 50$\mu \mathrm{g} / \mathrm{m}^3$ (poor) and 100$\mu \mathrm{g} / \mathrm{m}^3$ (very poor) at 548 locations distributed throughout the mainland of Spain, including a set of 0.5° $\times$ 0.5° (50km $\times$ 50km) grids (206 locations) and locations of all $\mathrm{PM}_{10}$ stations (342 monitors).

\begin{figure*} [ht]
	\centering
	\subfloat[\label{fig6:a}]{
		\includegraphics[scale=0.35]{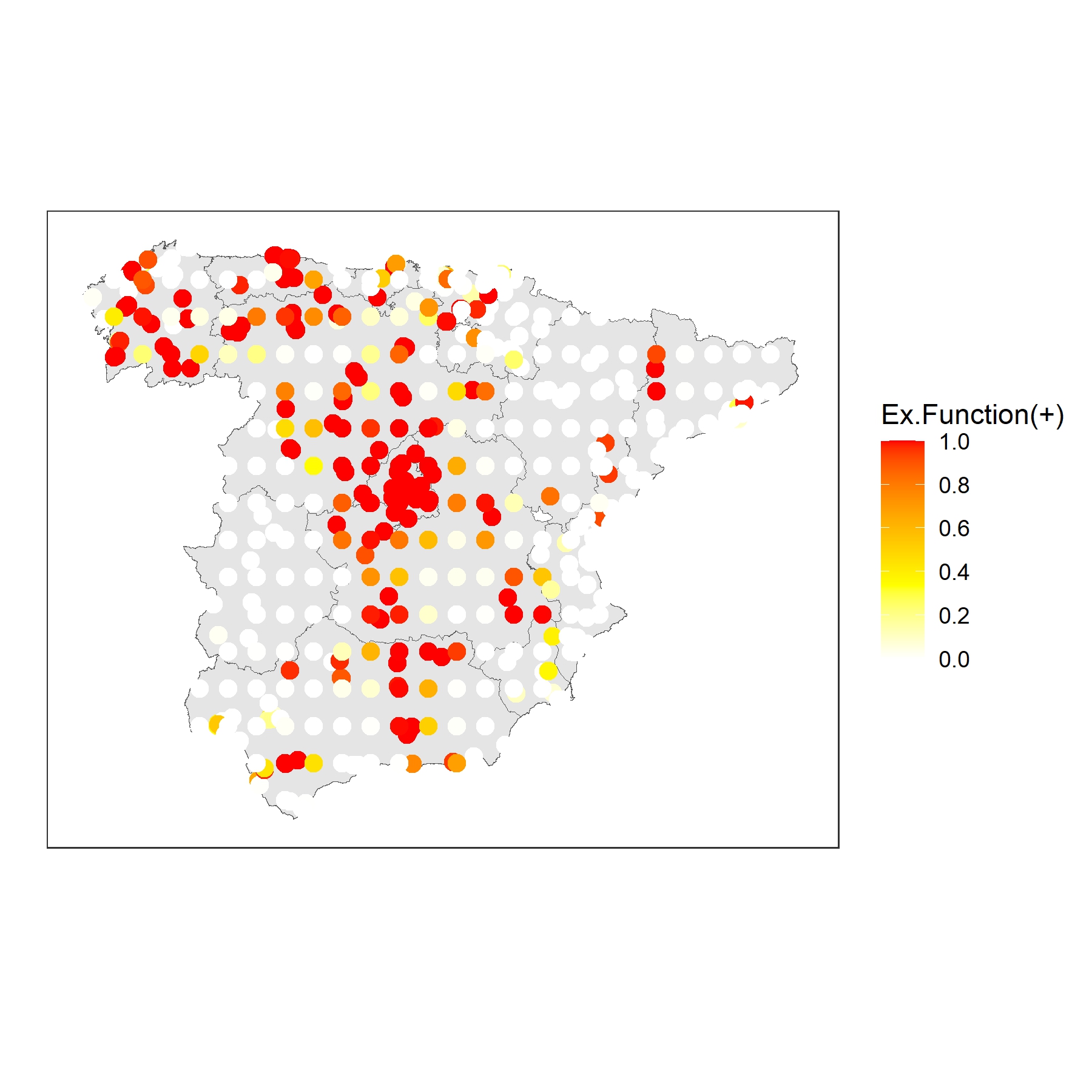}}
	\subfloat[\label{fig6:b}]{
		\includegraphics[scale=0.35]{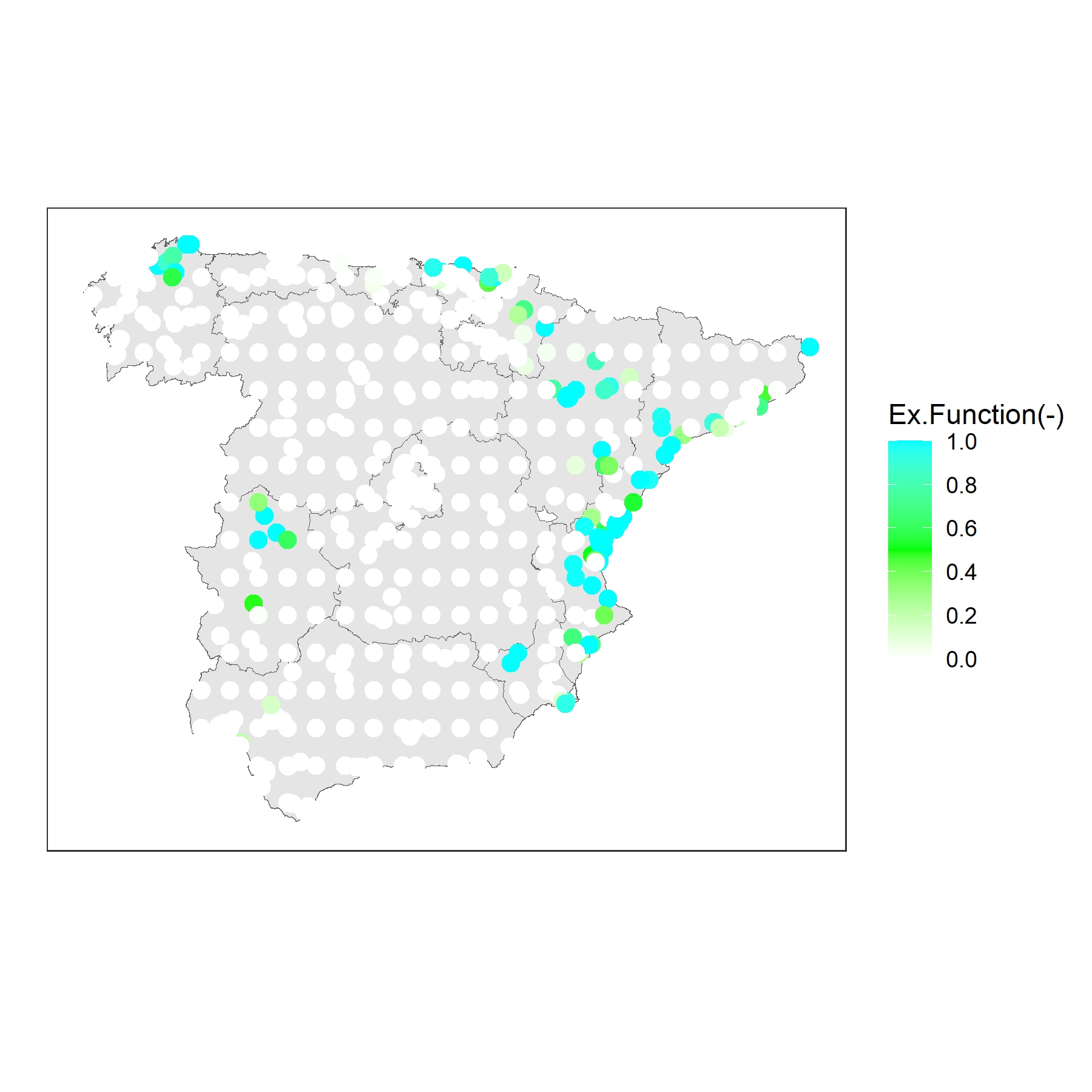}}
	\caption{Positive and negative excursion functions with threshold 50$\mu \mathrm{g} / \mathrm{m}^3$ are displayed in (a) and (b), respectively. Annual maximum concentration levels for locations in \textcolor{red}{red} are likely to exceed 50$\mu \mathrm{g} / \mathrm{m}^3$, and concentration levels for locations in \textcolor{cyan}{cyan} are probably below the threshold.}
	\label{EF50} 
\end{figure*}

\begin{figure*} [ht]
	\centering
	\subfloat[\label{fig7:a}]{
		\includegraphics[scale=0.35]{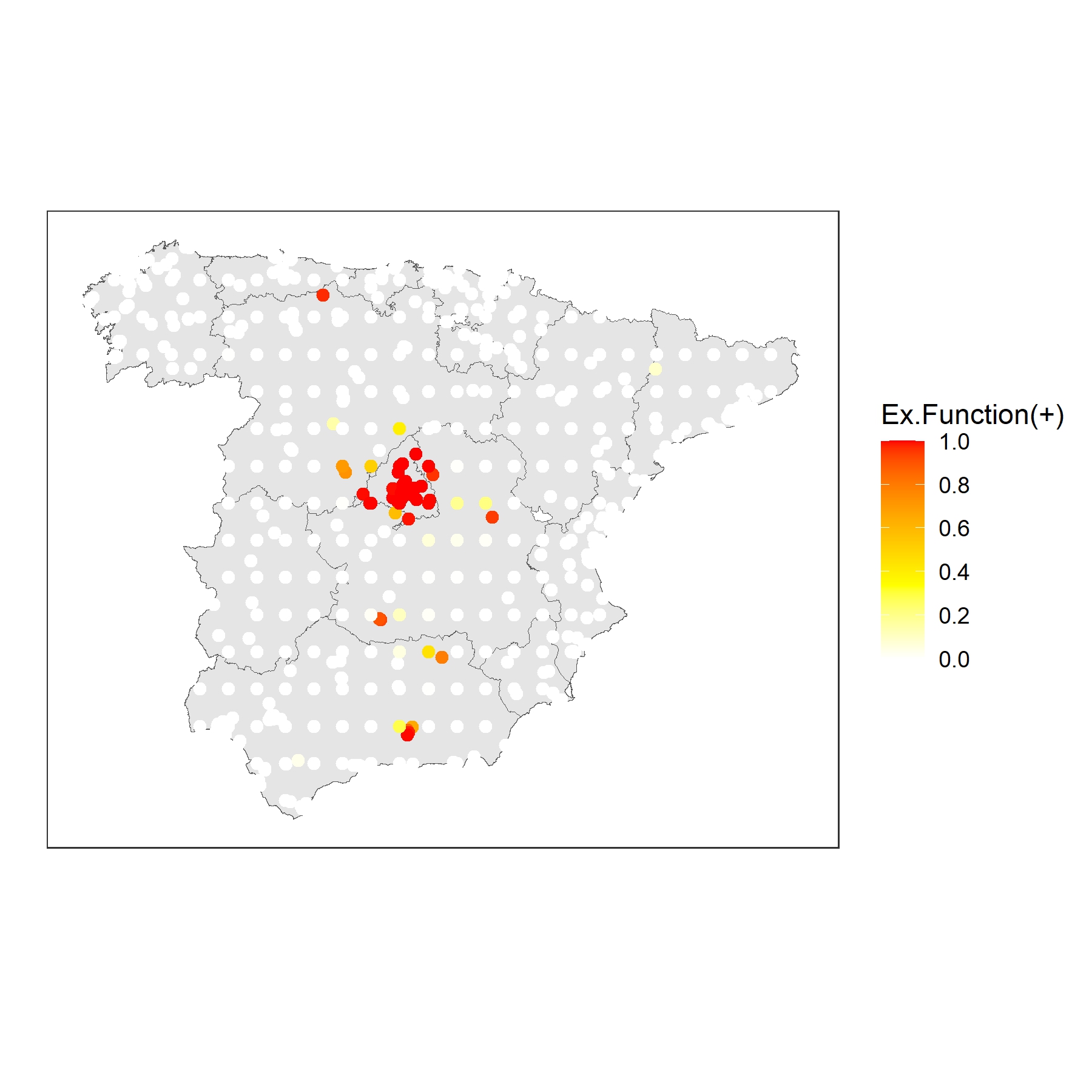}}
	\subfloat[\label{fig7:b}]{
		\includegraphics[scale=0.35]{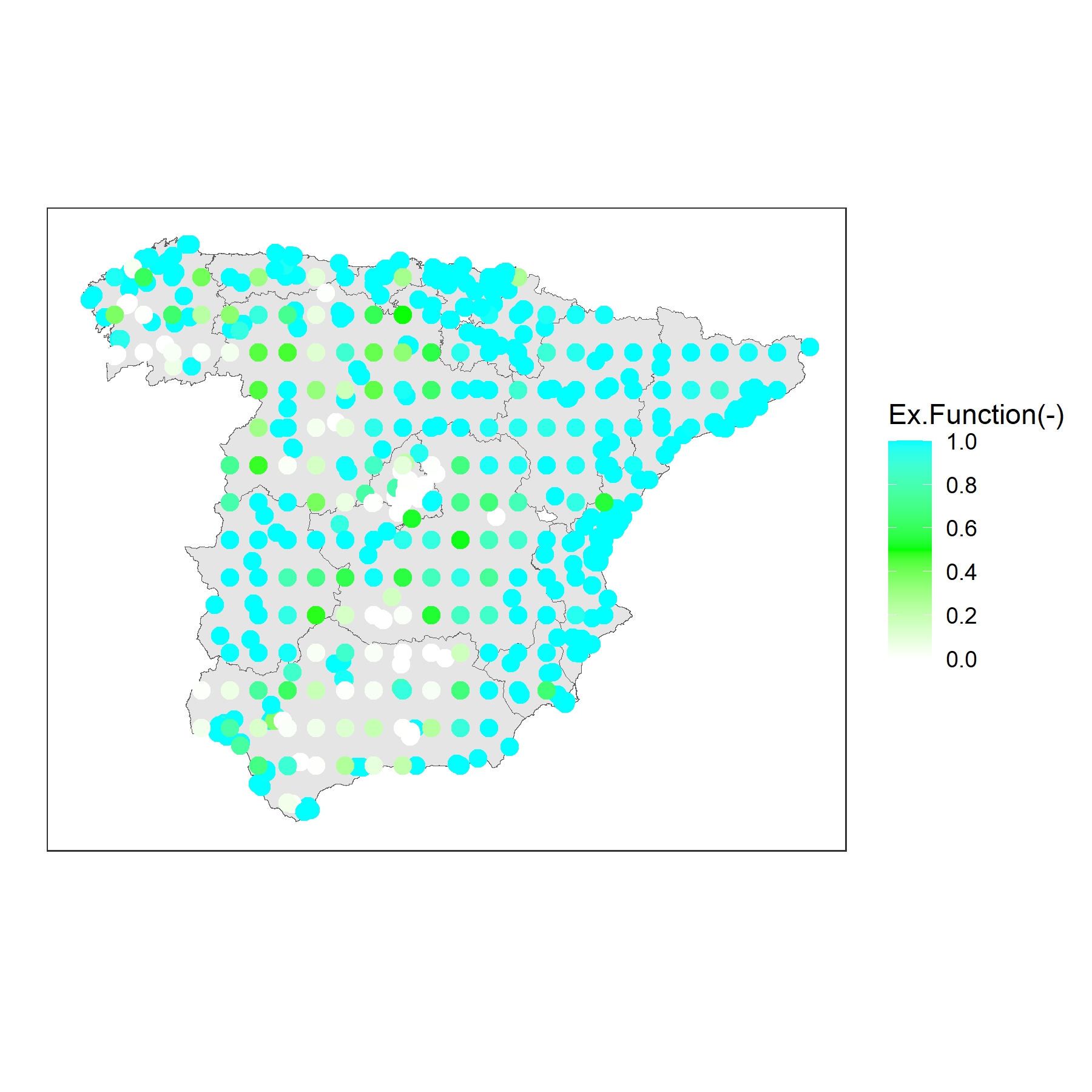}}
	\caption{ Positive and negative excursion functions with threshold 100$\mu \mathrm{g} / \mathrm{m}^3$ are displayed in (a) and (b), respectively. Annual maximum concentration levels for locations in \textcolor{red}{red} are likely to exceed 100$\mu \mathrm{g} / \mathrm{m}^3$, and concentration levels for locations in \textcolor{cyan}{cyan}  are probably below the threshold.}
	\label{EF100} 
\end{figure*}

In the case of exceeding 50$\mu \mathrm{g} / \mathrm{m}^3$ (Figure \ref{EF50}), the probability for simultaneously exceeding is high in the northwest, middle, and south, meaning that poor $\mathrm{PM}_{10}$ pollution probably hazard these locations during the year. In contrast, the negative excursion function with threshold 50$\mu \mathrm{g} / \mathrm{m}^3$ indicates that the regions in the north and east enjoy good or moderate air quality throughout the whole year. In the case of 100$\mu \mathrm{g} / \mathrm{m}^3$ (Figure \ref{EF100}), the probabilities for very poor $\mathrm{PM}_{10}$ pollution occurrence are low (in white) in most regions, but still likely to appear in certain areas in the community of Madrid, meanwhile, most areas in the north, northeast and east are expected to be below this threshold.

\section{Discussions and Conclusions}\label{conclusion}

In this paper, we discovered the spatio-temporal patterns of $\mathrm{PM}_{10}$ concentrations levels in mainland Spain under the framework of INLA-EVT methodology. The high-quality data-set of annual maxima and annual mean of daily PM$_{10}$ concentration levels was jointly studied successfully, due to the following three reasons. Firstly, Spain is a mountainous country with a large central plateau, and this complicated orography is usually associated with a variety of climatic conditions. Secondly, the air pollution monitors distributed throughout mainland Spain provide high-quality time series data, which supports both local and national level accurate prediction and credible inference. Finally, Spain generally enjoys good air quality  (low annual mean), but extreme air pollution does appear on certain days during the year (high annual maxima). This circumstance allows us to investigate the potential difference in the generation and spread of the moderate and extreme cases.

We establish a series of Bayesian spatio-temporal models on extreme $\mathrm{PM}_{10}$ concentrations in Spain, with specifically meteorological predictors, human effects, and spatio-temporal random effects following SPDE and AR(1) dynamic to account for the dependence not explained by covariates. We also provide evidence of similar and reverse connections between influential predictors and different scaled $\mathrm{PM}_{10}$ concentrations by the Bayesian joint model with sharing effects, as well as generate the annual maxima excursion functions maps specified at the grid level to highlight the regional risk ranking.

Although most statistical studies focus on moderate cases with long-term exposure, in the epidemiological field, short-term exposure to severe particulate matter is considered as an essential public health issue with major acute cardiovascular problems and health economic consequences. For example, \cite{BJextreme} {emphasized} that people living in highly polluted regions probably increase heart failure hospitalisations and cardiovascular mortality several folds. \cite{SHAH20131039} also {pointed} out that even modest improvements in air quality are projected to have major population health benefits and substantial health-care cost savings, preventing thousands of heart failure hospitalisations and saving millions of US dollars a year. 

Combined with these health and economic hazards, the main findings in this paper are expected to provide in-depth knowledge of extreme air pollution spreading, promote awareness of extreme value studies, and provide suggestions to the national governments regarding the legislation of extremely poor air quality regulation and human health protection. {In particular, the joint model incorporating sharing effects emphasizes the potential reverse or similar impacts of altitude and {precipitation} 
on both moderate and extreme cases of $\mathrm{PM}_{10}$ pollution. } Moreover, the excursion functions maps indicate that the central  region in Spain is more likely to experience severe $\mathrm{PM}_{10}$ pollution, which can be applied in research of long-term effects and health outcomes in epidemiological studies, such as acute cardiovascular events \citep{jama, SHAH20131039} and various types of strokes \citep{shortterm}.

In conclusion, our study provides valuable insights into the generation and spreading of extreme PM$_{10}$ pollution through innovative methods incorporating sharing effects in the joint model. This approach holds promise for future environmental and epidemiological studies in exploring various air pollutants and their relationship with meteorological variables and anthropogenic factors.

%% Loading bibliography style file
% \bibliographystyle{model1-num-names}
\bibliographystyle{apalike}

% Loading bibliography database
\bibliography{reference}

%\vskip3pt

\end{document}